\documentclass[a4paper,11pt]{article}
\pdfoutput=1 

\usepackage{jheppub} 
\usepackage{tikz}
\usepackage{hyperref}
\usepackage{cleveref}
\usepackage{supertabular}
\usepackage{todonotes}
\usepackage{mathrsfs}
\usepackage{array}
\usepackage{lipsum}
\usepackage{breqn}
\usepackage{comment}

\usepackage[T1]{fontenc} 

\newcommand{\sh}{\mathrm{sh}}
\newcommand{\ch}{\mathrm{ch}}

\newcommand{\Ch}{\mathrm{Ch}}

\def\q{{\rm q}}

\newcommand{\be}{\begin{equation}}
\newcommand{\ee}{\end{equation}}
\newcommand{\lt}{\left}
\newcommand{\rt}{\right}

\title{\boldmath More on 5d Wilson Loops in Higher-Rank Theories and Blowup Equations}

\author[1]{Minhao Liu,}
\author[2,3]{Xin Wang,}
\author[1,4]{and Rui-Dong Zhu}

\affiliation[1]{Institute for Advanced Study \& School of Physical Science and Technology,\\ Soochow University, Suzhou 215006, China}
\affiliation[2]{Interdisciplinary Center for Theoretical Study, University of Science and Technology of China, Hefei, Anhui 230026, China}
\affiliation[3]{Peng Huanwu Center for Fundamental Theory, Hefei, Anhui 230026, China}
\affiliation[4]{Jiangsu Key Laboratory of Frontier Material Physics and Devices,\\ Soochow University, Suzhou 215006, China}

\emailAdd{min-haoliu@outlook.com}
\emailAdd{wxin@ustc.edu.cn}
\emailAdd{rdzhu@suda.edu.cn}

\preprint{USTC-ICTS/PCFT-26-13}

\abstract{
In this article, we further explore the construction and computation of expectation values for Wilson loops in higher-rank 5d ${\cal N}=1$ gauge theories on $\mathbb{C}^2\times S^1$, by explicitly computing the Wilson loops via Chern-character insertion and qq-characters, including cases with the exceptional gauge group $G_2$.
In particular, we propose a systematic way to write down the general blowup equations for Wilson loops by using the constraints from the one-form symmetry and low-instanton data from the instanton partition function. In addition, for one-instanton contributions in a large family of Wilson loop representations, we observe that they admit a $q_1q_2$-expansion, similar to the Hilbert-series structure of instanton partitions in pure gauge theories.
}

\begin{document} 
\maketitle
\flushbottom

\section{Introduction}
\label{sec:intro}
Wilson loop operators are fundamental gauge-invariant, non-local operators in quantum field theory, and their vacuum expectation values (VEVs) provide essential probes of strongly coupled dynamics such as quark confinement. In non-supersymmetric theories such as QCD, the computation of Wilson loop VEVs remains notoriously difficult, with progress relying largely on lattice simulations and approximate techniques. In contrast, supersymmetric gauge theories offer a unique setting in which exact results can be computed. Supersymmetry protects certain observables, and the advent of localization techniques has enabled the non-perturbative evaluation of Wilson loops by reducing their path integrals to finite-dimensional matrix models \cite{Pestun:2007rz} (see also the review \cite{Pestun:2016zxk}). This has led to remarkable progress across various dimensions, from three-dimensional Chern–Simons–matter theories to four-dimensional $\mathcal{N}=4$ super Yang–Mills, where Wilson loops provide exact checks of holography, dualities, and integrability.

In five-dimensional supersymmetric gauge theories, Wilson loops acquire an even richer structure, closely tied to the dynamics of BPS states and string-theoretic constructions. Recent developments have demonstrated that, in the weak-gravity limit of five-dimensional $\mathcal{N}=1$ supergravity, the BPS partition function admits an expansion in terms of the VEVs of half-BPS Wilson loops in all representations of the gauge group in the emergent gauge theory \cite{Huang:2025xkc}. This remarkable observation underscores the importance of studying Wilson loops in five dimensions, particularly for higher-rank gauge groups and higher representations, where systematic results are still scarce.

The natural setting for investigating these observables is the $\Omega$-background $S^1 \times \mathbb{C}^2_{\epsilon_1,\epsilon_2}$, characterized by equivariant deformation parameters $\epsilon_{1,2}$. In this background, the BPS partition function can be expressed as an instanton expansion, with each instanton contribution identified with the Witten index of the ADHM gauged quantum mechanics. This framework admits a natural enrichment by introducing $\tfrac{1}{2}$-BPS line defects, in particular Wilson line operators extended along the temporal circle $S^1$. The presence of such defects modifies the instanton calculus, enriching the structure of the partition function and providing richer information about the BPS spectrum \cite{Nekrasov:2015wsu,Kimura:2015rgi,Bourgine:2015szm,Kim:2016qqs,Bourgine:2016vsq}.

Wilson loop operators carry charges under one-form symmetries, which have received increasing attention in recent years \cite{Gaiotto:2014kfa} (see also the lecture notes \cite{Schafer-Nameki:2023jdn,Luo:2023ive}). From the perspective of M-theory, half-BPS Wilson loops in a five-dimensional gauge theory engineered by a Calabi–Yau threefold $X$ can be realized as insertions of M2-branes wrapping relative 2-cycles in $X$ \cite{Kim:2021gyj}. This construction naturally leads to a refined BPS expansion for Wilson loop expectation values, generalizing the celebrated Gopakumar–Vafa expansion for closed topological strings \cite{Gopakumar:1998ii,Gopakumar:1998jq}. The refined expansion with Wilson loop insertions provides a powerful bridge between gauge theory observables, enumerative geometry, and string theory. However, explicit studies so far have largely focused on rank-one gauge groups and low-dimensional representations, leaving the {\it higher-rank} and {\it higher-representation} cases largely unexplored. One of the central aims of this paper is to fill this gap.

Another important structural ingredient is provided by the blowup equations \cite{Nakajima:2003pg,Nakajima:2005fg,Nakajima:2009qjc}, a set of functional relations satisfied by BPS partition functions of four-, five- and six-dimensional supersymmetric gauge theories \cite{Huang:2017mis,Gu:2017ccq,Gu:2018gmy,Bershtein:2018zcz,Gu:2019dan,Kim:2019uqw,Gu:2019pqj,Shchechkin:2020ryb,Gu:2020fem,Bershtein:2013oka}. These equations encode profound constraints on partition functions and have proved extremely powerful in computing refined BPS invariants. Recently, the blowup equations have been generalized in the presence of Wilson loop operator insertions \cite{Kim:2021gyj,Wang:2023zcb}. One goal of this work is to further study blowup equations with Wilson-loop insertions. More precisely, we provide a systematic prescription to formulate blowup equations. 

In this work, we investigate these interrelated themes in the setting of five-dimensional $\mathcal{N}=1$ supersymmetric gauge theories of higher rank. Our aim is twofold. First, we apply and extend equivariant Chern-character insertions and qq-character methods to compute Wilson-loop expectation values in a wide range of representations, and across both classical gauge groups and exceptional examples such as $G_2$. This enables tests of refined BPS expansions across a broader range of gauge groups and representations than previously analyzed. By using these results, we observe that the one-instanton contributions to a large family of Wilson loop VEVs admit a {\it Hilbert-series-like expansion}, as shown in Eq. \eqref{eq:1-ins-Wilson2} and Eq. \eqref{1-ins-Wilson}.
Second, we further explore the formulation of blowup equations in the presence of Wilson-loop insertions, emphasizing how {\it one-form symmetry} constrains their structure and how the remaining coefficients can be fixed from low-instanton data. Together, these results provide a coherent framework for analyzing Wilson-loop observables, elucidate their role in refined BPS spectra, and reveal structural connections to instanton geometry and topological string theory.

The article is organized as follows. In Section~\ref{sec:5dwilson}, we review the basic properties of Wilson loops that are relevant to our paper. In Section~\ref{sec:Computation}, we provide examples and detailed calculations of Wilson loop VEVs for higher-rank gauge groups and present the Hilbert-series-like structure for the one-instanton BPS sectors for Wilson loops.
In Section~\ref{s:blowup}, we demonstrate how one-form symmetry can be used to constrain the blowup equation in the presence of Wilson loops and provide explicit checks for the generalized blowup equation with Wilson loops. In Section~\ref{s:sum}, we provide a summary and future directions.

\section{Wilson loops in 5d ${\cal N}=1$ theories}\label{sec:5dwilson}

\begin{figure}
\centering
\begin{tikzpicture}[x=0.75pt,y=0.75pt,yscale=-1,xscale=1]
\draw   (251,101) -- (360,101) -- (360,210) -- (251,210) -- cycle ;
\draw    (211,61) -- (251,101) ;
\draw    (360,210) -- (400,250) ;
\draw [color={rgb, 255:red, 74; green, 144; blue, 226 }  ,draw opacity=1 ][line width=2.25]    (401,60) -- (360,101) ;
\draw    (251,210) -- (210,251) ;
\draw    (401,60) -- (449,60) ;
\draw    (401,60) -- (401,22) ;
\draw    (406.91,74.16) -- (382.09,99.84) ;
\draw [shift={(380,102)}, rotate = 314.03] [fill={rgb, 255:red, 0; green, 0; blue, 0 }  ][line width=0.08]  [draw opacity=0] (8.93,-4.29) -- (0,0) -- (8.93,4.29) -- cycle    ;
\draw [shift={(409,72)}, rotate = 134.03] [fill={rgb, 255:red, 0; green, 0; blue, 0 }  ][line width=0.08]  [draw opacity=0] (8.93,-4.29) -- (0,0) -- (8.93,4.29) -- cycle    ;
\draw (407,88) node [anchor=north west][inner sep=0.75pt]   [align=left] {$\displaystyle m\ \rightarrow \infty $};
\end{tikzpicture}
    \caption{Adding flavor branes and taking the mass parameters to infinity, one can realize the Wilson loops in tensor product of fundamental or anti-fundamental representations and read off their contributions from the asymptotic behavior. }
    \label{fig:wilson-brane}
\end{figure}
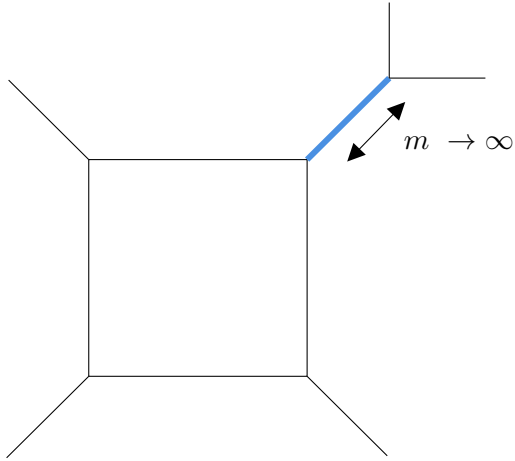

In this article, we consider 5d ${\cal N}=1$ gauge theories on a product manifold of a circle and the $\Omega$-background, $S^1\times \mathbb{C}^2_{\epsilon_1,\epsilon_2}$. These theories can often be engineered
from M-theory compactification on a non-compact Calabi-Yau threefold $X$ \cite{Katz:1996fh} and the M-theory setup is dual to the type IIB string-theory construction based on $(p,q)$-brane webs \cite{Aharony:1997bh,Leung-Vafa}. One can compute the partition function of the gauge theory on the $\Omega$-background in an exact way, which is often called the Nekrasov partition function, by using the techniques developed in the topological string theory on the Calabi-Yau $X$ \cite{Aganagic:2003db,Iqbal:2007ii} or the localization method for supersymmetric theories \cite{Nekrasov:2002qd,Tachikawa:2004ur}. The partition function encodes the degeneracy of BPS particles, $N^\beta_{j_L,j_R}$, with charge $\beta\in H_2(X,\mathbb{Z})$ and $(j_L,j_R)$ labeling the representation in the 5d little group SU(2)$_L\times$SU(2)$_R$. One can insert $\frac{1}{2}$-BPS Wilson loops wrapping the $S^1$-direction (and locating at the origin of $\mathbb{C}^2_{\epsilon_1,\epsilon_2}$) into the theory, 
\begin{equation}
    W_{\bf r}={\rm tr}_{\bf r}\,T\exp\lt(i\oint_{S^1}{\rm d}t\lt[A_0(t)-\phi(t)\rt]\rt),
\end{equation}
where $T$ denotes the time ordering, and it is still possible to compute the expectation values of such Wilson loops, including the instanton corrections, with the localization technique \cite{Tong:2014cha,Gaiotto:2015una,Kim:2016qqs}. At the level of concrete computation, brane constructions help to compute the Wilson loops. In this work, we mainly rely on two of them. The first one directly introduces a Wilson loop. In the M-theory picture, a Wilson loop can be realized by inserting a heavy stationary quark coming from an M2-brane wrapping a non-compact holomorphic 2-cycle in the Calabi-Yau geometry \cite{Kim:2021gyj}. It shares the same idea as the brane realization of Wilson lines in the holography context \cite{Maldacena:1998im,Rey:1998ik,Gomis:2006sb,Tong:2014cha} with semi-infinite F1 strings stretching between (inserted) D3 and (color) D5-branes. An alternative realization in the same spirit was proposed in the 5-brane web in \cite{Huang:2022hdo} to generate the $\frac{1}{2}$-BPS Wilson loops from gauge theories with additional hypermultiplets. One can add D5 flavor branes to the brane web of the original gauge theory, and take the large mass limit to extract Wilson loops in the tensor-product representation of the fundamental or anti-fundamental representations of SU($N$) (see Figure \ref{fig:wilson-brane}). This directly inspires a systematic way to compute the Wilson loops by inserting Chern characters into the ADHM construction, and we give the details of this method in Section \ref{s:Chern}. The second approach to introduce a heavy quark is based on the type IIA picture by adding D4${}^{\prime}$ branes wrapping the $S^1$-direction and having the other 4 directions perpendicular to the color D4 branes (see Table \ref{t:qq-brane}). This gives rise to a family of line defects called qq-characters \cite{Tong:2014cha,Nekrasov:2015wsu,Kim:2016qqs}. It again allows us to compute such line defects via the ADHM construction and the localization. The qq-character gives a linear combination of different Wilson loops. For example, when we insert only one D4${}^{\prime}$ brane, we call the line defect a fundamental qq-character, and it is given by the sum of antisymmetric representations, 
\begin{equation}\label{qq-Wilson}
\chi^G(x)=x^{-\frac{\dim \square}{2}}\sum_{k=0}^{\dim \square}(-x)^{k}Z_{W_{\wedge^k\square}},
\end{equation}
where $\bigwedge^k\square$ denotes the $k$-th antisymmetric tensor product of the fundamental/vector representation denoted by $\square$ of the gauge group $G$. $x=e^{-m_{\rm D4'}}$ denotes the position of $\rm D4'$ branes in the 9-direction. 

\begin{table}
\begin{center}
\begin{tabular}{|c|c|c|c|c|c|c|c|c|c|c|}
\hline
& 0 & 1 & 2 & 3 & 4& 5 & 6 & 7 & 8 & 9 \cr
\hline
D4/O4 & $\bullet$ & $\bullet$ & $\bullet$ & $\bullet$ & $\bullet$ & $-$ & $-$ & $-$ & $-$ & $-$ \cr
\hline
D0 & $\bullet$ & $-$ & $-$ & $-$ & $-$ & $-$ & $-$ & $-$ & $-$ & $-$ \cr
\hline
D$4'$ & $\bullet$ & $-$ & $-$ & $-$ & $-$ & $\bullet$  & $\bullet$ & $\bullet$ & $\bullet$ & $-$ \cr
\hline
\end{tabular}
\caption{Brane web construction of the ADHM construction for qq-characters (introduced by D4' branes). }
\label{t:qq-brane}
\end{center}
\end{table}

As a generalization of the refined BPS expansion of the Nekrasov partition function \cite{Gopakumar:1998jq,Gopakumar:1998ii}, it was conjectured in \cite{Huang:2022hdo,Kim:2021gyj} that an analogous BPS expansion, with non-negative expansion coefficients, can also be applied to Wilson loops. 
On the Coulomb branch of a rank-$r$ 5d $\mathcal{N}=1$ theory, the real scalar in the vector multiplet has non-trivial VEVs denoted by $a_i,i=1,\cdots,r$. The gauge group $G$ is broken to its Abelian subgroup $U(1)^r$ and Wilson-loop representations can be labeled by their $U(1)$ charges $[{\rm q}_1,\cdots,{\rm q}_r]$. The corresponding expectation values of the Wilson loops can be written as a sum of BPS sectors $\mathcal{F}_{[i_1,\cdots,i_r]}$ as 
\begin{align}
    \left\langle W_{[{\rm q}_1,\cdots,{\rm q}_r]}\right\rangle =\left(\prod_{j=1}^{r}{\rm q}_j!\right)\cdot\sum_{\ell=1}^{|{\rm q}|}\sum_{\substack{n_1+\cdots +n_k=\ell\\n_1 {\rm q}_{j,1}+\cdots +n_k{\rm q}_{j,k}={\rm q}_j;\\n_i,{\rm q}_{j,i}\geq 0} }\frac{{n_1}!\cdots n_k!}{(n_1+\cdots +n_k)!}\prod_{i=1}^k\frac{\mathcal{F}^{n_i}_{[{\rm q}_{1,i},\cdots,{\rm q}_{r,i}]}}{({\rm q}_{1,i}!\cdots {\rm q}_{r,i}!)^{n_i}}.
\end{align}
When the 5d theory has a gauge theory description, the charge vector indicates that the Wilson loop is in the tensor product representation $\mathbf{r}=\bigotimes_{i=1}^r\mathbf{r}_i^{\otimes \q_i}$, where $\mathbf{r}_i$ denotes the $i$-th fundamental representation\footnote{The $i$-th fundamental representation is defined as the representation whose highest weight is the $i$-th fundamental weight $\omega_i=[0,\cdots,0,1_{i\text{-th}},0,\cdots,0]$.} of the gauge group. The BPS sector $\mathcal{F}_{[\q_1,\cdots,\q_r]}$ admits the expansion
\begin{align}
    \mathcal{F}_{[{\rm q}_1,\cdots,{\rm q}_r]}=\mathcal{I}^{|{\rm q}|-1}\,e^{{\rm q}\cdot a}\,\sum_{\beta,j_L,j_R}(-1)^{2j_L+2j_R}\widetilde{N}^{\beta}_{j_L,j_R}\chi_{j_L}(\epsilon_-)\chi_{j_R}(\epsilon_+)e^{-\beta\cdot t},\quad \text{if }|{\rm q}|>0,
\end{align}
where the factor $e^{{\rm q}\cdot a}$ is the perturbative contribution from the charges of Wilson loops, and the non-negative integers $\widetilde{N}^{\beta}_{j_L,j_R}$ are the refined BPS invariants for Wilson loops, which count the BPS states of M2-branes wrapping non-compact primitive 2-cycles in the M-theory picture. $\mathcal{I}$ is the momentum factor 
\begin{align}
    \mathcal{I}=2\sinh{\frac{\epsilon_1}{2}}\cdot 2\sinh{\frac{\epsilon_2}{2}}.\label{def-I}
\end{align}
Here we also used the notation $|{\rm q}|={\rm q}_1+\cdots+{\rm q}_r$, $\epsilon_{\pm}=\frac{1}{2}(\epsilon_1\pm\epsilon_2)$ and
\begin{align}
    \chi_j(\epsilon):=\sum_{k=-j}^j{e^{k\epsilon}}.
\end{align}
For example, in the rank-2 gauge theory, the expectation values of Wilson loops have the expansions:
\begin{align}
    \left\langle W_{[1,0]}\right\rangle ={\mathcal{F}_{[1,0]}},\quad \left\langle W_{[0,1]}\right\rangle ={\mathcal{F}_{[0,1]}},
\end{align}
\begin{align}
    \left\langle W_{[1,1]}\right\rangle =\mathcal{F}_{[1,0]}\mathcal{F}_{[0,1]}+\mathcal{F}_{[1,1]},\quad \left\langle W_{[2,0]}\right\rangle =\mathcal{F}_{[1,0]}^2+\mathcal{F}_{[2,0]},\quad \left\langle W_{[0,2]}\right\rangle =\mathcal{F}_{[0,1]}^2+\mathcal{F}_{[0,2]},
\end{align}
\begin{align}
    \left\langle W_{[3,0]}\right\rangle &=\mathcal{F}_{[1,0]}^3+3\mathcal{F}_{[1,0]}\mathcal{F}_{[2,0]}+\mathcal{F}_{[3,0]}, \\
    \left\langle W_{[2,1]}\right\rangle &=\mathcal{F}_{[1,0]}^2\mathcal{F}_{[0,1]}+\mathcal{F}_{[2,0]}\mathcal{F}_{[0,1]} +2\mathcal{F}_{[1,0]}\mathcal{F}_{[1,1]}+\mathcal{F}_{[2,1]},\\
    & \qquad\qquad\qquad\qquad\vdots \nonumber
\end{align}
The non-negative integer nature of $\widetilde{N}^{\beta}_{j_L,j_R}$ provides consistency checks for our computation of Wilson loops. 

Higher-form symmetries act on extended objects/defects \cite{Gaiotto:2014kfa}, and a $p$-dimensional defect in $d$-dimensional QFT can potentially carry a $p$-form symmetry charge, generated by a topological operator of $d-1-p$ dimensions. In the case of Wilson loops in gauge theory with gauge group $G$, they transform under the one-form symmetry corresponding to the center of $G$. One may refer to e.g. Table 1 in \cite{Bhardwaj:2023kri} for the one-form symmetry and the charge of Wilson loops in fundamental representations. For later convenience, let us briefly describe the one-form symmetry of $A_n$ and $B_n$ gauge groups. For $A_n$-type gauge groups, the one-form symmetry is $\mathbb{Z}_{n+1}$ and the charge of a Wilson loop in representation with weight $\omega=\sum_{i=1}^n\mu_i\omega_i$ is given by 
\begin{equation}
    c_{\bf r}=\sum_ii\mu_i\mod n+1, 
\end{equation}
where $\omega_i$ stands for the $i$-th fundamental weight. For $B_n$- and $C_n$-type gauge groups, the one-form symmetry is always $\mathbb{Z}_2$. In the case of $B_n$-type gauge groups, it is simple to compute the charge of Wilson loops, as only the weight of the spinor representation has a non-trivial charge of $\mathbb{Z}_2$ and the charge of a general representation is determined by the number of fundamental weights of spinor contained. One may refer to \cite{Morrison:2020ool} for more details on the one-form and two-group symmetries in 5d ${\cal N}=1$ theories.

One motivation for this article is to further explore a set of constraint equations, called blowup equations, for Wilson loops in arbitrary representations in higher-rank gauge theories. Such equations were originally proposed in \cite{Nakajima:2003pg,Nakajima:2005fg,Nakajima:2009qjc} for the 4d/5d BPS partition functions of certain theories in the form,
\begin{equation}
    \hat{{\cal Z}}=\sum_{\vec{n}}{\cal Z}^{({\rm N})}(\vec{n}){\cal Z}^{({\rm S})}(\vec{n}),
\end{equation}
where $\hat{{\cal Z}}$ denotes the full partition function on the $\Omega$-background with one point blowup, and it can alternatively be given by the contributions from the north pole ${\cal Z}^{({\rm N})}$ and south pole ${\cal Z}^{({\rm S})}$ of the blowup $\mathbb{P}^1$. $\vec{n}$ are the magnetic fluxes on $\mathbb{P}^1$. The blowup equations were then developed in \cite{Kim:2021gyj,Wang:2023zcb} as constraints on Wilson loops in simple cases. It is natural to expect that the one-form symmetry imposes some restrictions on the blowup equation of Wilson loops, and this constitutes a central result of this work. In Section \ref{s:blowup}, we present a systematic algorithm to fix the blowup equations, i.e. $\hat{{\cal Z}}$ as a linear combination of Wilson loops, with the information from the first few BPS spectrum data plus the one-form symmetry constraints.

Let $v:=\sqrt{q_1q_2}$ and $x:=\sqrt{q_1/q_2}$. The one-instanton part of BPS sector $\mathcal{F}^{(1)}_{[{\rm q}_1,\cdots,{\rm q}_r]}$, defined from the instanton expansion,
\begin{align}
    \mathcal{F}_{[{\rm q}_1,\cdots,{\rm q}_r]}=\sum_{k=0}^{\infty}{\frak{q}}^k\,\mathcal{F}^{(k)}_{[{\rm q}_1,\cdots,{\rm q}_r]}\,,
\end{align}
has the $v$-expansion
\begin{equation}\label{eq:1-ins-Wilson2}
    \mathcal{F}_{[{\rm q}_1,\cdots,{\rm q}_r]}={\cal I}^{|{\rm q}|-1}\sum_{i=0}^{\infty}f_i(\chi_{{\bf r}^{\prime}},\chi_j(\epsilon_-))v^{i},
\end{equation}
where 
\begin{align}
    f_i(\chi_{{\bf r}^{\prime}},\chi_j(\epsilon_-))=\sum_{{\bf r}^{\prime},j}c_{{\bf r}^{\prime},j}\,\chi_{{\bf r}^{\prime}}\,\chi_j(\epsilon_-).
\end{align}
Here $\chi_{{\bf r}^{\prime}}$ is a character of the gauge group with the same one-form symmetry charge as ${\bf r}$ and $\chi_j(\epsilon_-)$, as a Laurent polynomial of $x$, is a spin-$j$ character for the SU(2)$_L$ symmetry in the 5d little group for massive particles. An interesting observation in this work is that the one-instanton part of BPS sector $\mathcal{F}^{(1)}_{[{\rm q}_1,\cdots,{\rm q}_r]}$ for a large family of Wilson loops,  up to the factor ${\cal I}^{|{\rm q}|-1}$, does not depend on $x=\sqrt{q_1/q_2}$.
For example, consider the Wilson loop in the $i$-th fundamental representation ${\bf r}_i$ of SU($N$) gauge group, 
\begin{equation}
    \mathcal{F}^{(1)}_{{\bf r}_i}=-\sum_{k=0}^{\infty}\chi_{[k,0,\cdots,0,1_{i\text{-th}},0,\cdots,0,k]}v^{2i}.\label{1-ins-Wilson}
\end{equation}
Here $\chi_{[k,0,\cdots,0,1_{i\text{-th}},0,\cdots,0,k]}$ represents the character of the representation whose highest weight is $[k,0,\cdots,0,1_{i\text{-th}},0,\cdots,0,k]$.
We observe here a similar structure in Wilson loops to that observed in the relation between the Nekrasov partition function without matter and its corresponding Hilbert series \cite{Benvenuti:2010pq,Rodriguez-Gomez:2013dpa}.  The universality of this observation across different gauge groups and representations, along with the underlying rationale, presents an intriguing avenue for future exploration. We leave this investigation for subsequent work.

In the case that all Coulomb parameters are zero, we observe that the one-instanton BPS sector has a universal pattern
\begin{align}\label{eq:pattern}
     \mathcal{F}_{[{\rm q}_1,\cdots,{\rm q}_r]}|_{a\rightarrow 0}={\cal I}^{|{\rm q}|-1}\frac{g(v,x)}{(1-v^2)^{2h_G^{\vee}-2}},
\end{align}
where $g(v)$ is a palindromic polynomial of degree $2h_G^{\vee}-2$ that has the same pattern as the unrefined Hilbert series studied in \cite{Benvenuti:2010pq}. We have verified that the pattern \eqref{eq:pattern} works for all the examples we have studied in Section \ref{s:w-example} and in Section \ref{sec:4.1}.

\section{Computation of Wilson loops}\label{sec:Computation}

In this section, we review and present two efficient methods to compute the Wilson loops beyond the fundamental representations. The main idea for these methods is to insert static, heavy gauge charged particles. The statistical properties, e.g. Bosonic or Fermionic, provide two independent methods for the calculations. We apply these methods to compute the VEVs of Wilson loops for 5d pure $G_2$ theory. Furthermore, we compare the calculations for theories that have the isomorphic gauge algebra. The results obtained here are used to check the consistency of the blowup equation in Section \ref{s:blowup}. 

\subsection{Chern character insertion}\label{s:Chern}

The first approach is based on the picture provided in Figure \ref{fig:wilson-brane}, i.e. generating Wilson loops by inserting hypermultiplet contributions and taking the large-mass limit. For gauge theories with classical gauge groups, this is equivalent to inserting equivariant Chern characters in the ADHM integrand \cite{Losev:2003py,Shadchin:2004yx,Gaiotto:2015una}, 
\begin{equation}
    \Ch_{\bf F}(A_i,e^{-\phi_j},q_1,q_2)=\chi_{\rm fund}(A_i)-(1-q_1)(1-q_2)q_1^{-\frac{1}{2}}q_2^{-\frac{1}{2}}\hat{\chi}_{\rm fund}(e^{-\phi_j}),
    \label{chern-character}
\end{equation}
where $\chi_{\rm fund}$ stands for the character of the fundamental representation of the gauge group $G$, and $\hat{\chi}_{\rm fund}$ is the fundamental character of the dual gauge group $\hat{G}_k$ in the ADHM quantum mechanics. Chern characters associated to other representations can be constructed in terms of the above fundamental one, $\Ch_{\bf F}$ (see \cite[A.14]{Gaiotto:2015una}). For instance, for antisymmetric and anti-fundamental representations, we have
\begin{align}
    \Ch_{\bf \Lambda^2}(A_i,\phi_j,q_1,q_2)&=\frac{1}{2}\left(\Ch_{\bf F}(A_i,\phi_j,q_1,q_2)^2-\Ch_{\bf F}(A_i^2,\phi_j^2,q_1^2,q_2^2)\right)\\
    \Ch_{ \overline{\bf F}}(A_i,\phi_j,q_1,q_2)&=\Ch_{\bf F}(A_i^{-1},\phi_j^{-1},q_1^{-1},q_2^{-1})
\end{align}
More concretely, the $k$-instanton partition function with hypermultiplets in representations $R_i,\,i=1,\cdots,f$ is given by a contour integral:
\begin{equation}
    Z^{(k)}_{\text{pre}}=\frac{\mathfrak{q}^k}{|W(\hat{G}_k)|}\oint\prod_{j=1}^{k}\frac{{\rm d}\phi_j}{2\pi i} Z^{(k)}_{\textrm{vec}} \,{\rm PE}\left[\sum_{i=1}^f \frac{\Ch_{R_i}\,e^{-m_i}}{(1-q_1)(1-q_2)q_1^{-1/2}q_2^{-1/2}}\right],\label{Z-pre}
\end{equation}
where $Z^k_{\rm vec}$ is the contribution of the vector multiplet at $k$-instanton level. For the sake of being self-contained, we list the integrand coming from the vector multiplet well-known in the literature \cite{Moore:1997dj,Moore:1998et,Losev:1999nt,Losev:1999tu,Nekrasov:2002qd,Nekrasov:2004vw,Hwang:2014uwa}. In SU($N$) theories, it is given by \footnote{Here, we define $\sh(x)=e^{x/2}-e^{-x/2}.$}
\begin{equation}\label{eq:ZvecSUN}
    Z_{\text{vec}}^{(k)}=\prod_{I\neq J}^k \sh(\phi_I-\phi_J)\prod_{I,J=1}^k\frac{\sh(\phi_I-\phi_J+2\epsilon_+)}{\sh(\phi_I-\phi_J+\epsilon_+\pm\epsilon_-)}\prod_{I=1}^k\prod_{i=1}^N\frac1{\sh(\pm(\phi_I-a_i)+\epsilon_+)},
\end{equation}
and in SO($2N+\delta$) theories ($\delta=0,1$), 
\begin{align}
    Z_{\text{vec}}^{(k)}&=\left(\frac{\sh\left(\epsilon^+\right)}{\sh\left(\epsilon_1\right)\sh\left(\epsilon_2\right)}\right)^k\prod_{I=1}^{k}\frac{\sh\left(\pm2\phi_{I}\right)\sh\left(\pm2\phi_{I}+2\epsilon_{+}\right)}{\sh^\delta(\pm \phi_i+\epsilon_+)\prod_{i=1}^{N}\sh\left(\pm\phi_{I}\pm a_{i}+\epsilon_{+}\right)}\notag
    \\
    &\times\prod_{I<J}^{k}\frac{\sh\left(\pm\phi_{I}\pm\phi_{J}\right)\sh\left(\pm\phi_{I}\pm\phi_{J}+2\epsilon_{+}\right)}{\sh\left(\pm\phi_{I}\pm\phi_{J}\pm\epsilon_{-}+\epsilon_{+}\right)}.
\end{align}
The localization formula becomes much more involved for Sp($N$) theories, and the integrand is summarized in Appendix \ref{a:Sp}.

If the representations of the hypermultiplets are in the $i$-th fundamental representation of the gauge group, in the large mass limit of mass parameters, these hypermultiplets can be effectively treated as the sources for Wilson loops.
One can expand the pre-generating function \eqref{Z-pre} with respect to the effective mass $$M_i=\frac{e^{-m_i}}{(1-q_1)(1-q_2)q_1^{-1/2}q_2^{-1/2}}$$ to compute the VEVs for Wilson loops, but one should notice that there can potentially exist an overall extra factor $Z_{\rm extra}$ that needs to be removed s.t. we can define the following generating function, 
\begin{equation}
    Z_{\rm gen}:=Z^{-1}_{\rm extra}\sum_{k=0}^\infty Z^k_{\rm pre}=Z^G_{\rm inst}\left(1+\sum_i\left\langle W_{R_i}\right\rangle M_i+\sum_i\left\langle W_{R_i\otimes R_j}\right\rangle M_i M_j+\dots\right).
\end{equation}
It is then straightforward to read off the normalized expectation value of Wilson loops, $\left\langle W_R\right\rangle:=\frac{Z_{W_R}}{Z_{\rm inst}}$. The extra factor in general depends on Coulomb parameters; however, for a certain number of matter contents, $Z_{\rm extra}$ is 1 or independent of Coulomb parameters. One may refer to the examples shown in Section \ref{s:w-example} for more details.

To work out the extra factor in the case when the extra factor is independent of Coulomb parameters, one can consider the asymptotic behavior of the instanton partition function where all the Coulomb branch parameters are large, i.e., for any simple root $\alpha_i$ of $G$, $Q_{i}=e^{-\alpha_i\cdot a}\to 0$, 
\begin{equation}
    Z_{\rm extra}(m,{\frak{q}})=\left.Z_{\text{gen}}\right|_{ Q_{i}\to 0}. \label{extra-factor}
\end{equation}
The extra factor depends on the number and the representations of the hypermultiplets added, and roughly speaking, one can understand it as a certain contribution from parallel flavor branes in the IIB $(p,q)$-brane webs.

\subsection{qq-character}\label{s:qq}

Another indirect but systematic way to compute the Wilson loops is via the qq-characters \cite{Nekrasov:2015wsu}, whose ADHM brane construction was presented in \cite{Tong:2014cha} and is shown in  Table \ref{t:qq-brane}. The corresponding defect partition function can be calculated using the localization technique as \footnote{Within this subsection, the notation $x_j$ refers specifically to defect fugacities. This should not be confused with the parameter $x$, defined as $x = \sqrt{q_1/q_2}$, which appears in other sections.}
\begin{equation}
    Z_{\rm defect}(\{x\})=\sum_{k=0}^\infty\frac{\mathfrak{q}^k}{|W(\hat{G}_k)|}\oint_{C_{\textrm{JK}}}\prod_{I=1}^{k}\frac{{\rm d}\phi_I}{2\pi i} Z^{k}_{\textrm{vec}} Z^{(k)}_{\rm defect}(\{x\}),
\end{equation}
where in SU($N$) theories, $Z^{(k)}_{\rm defect}$ is given by \cite{Nekrasov:2015wsu,Kim:2016qqs}
\begin{equation}
    Z_{\text{defect}}^{(k)}(\{x_j\}_{j=1}^w)=\prod_{j=1}^w\prod_{i=1}^N \sh(X_j-a_i)\prod_{I=1}^k\frac{\sh(\pm(\phi_I-X_j)+\epsilon_-)}{\sh(\pm(\phi_I-X_j)-\epsilon_+)},\quad x_j=\exp(-X_j),\label{defect-SU}
\end{equation}
and in SO($2N+\delta$) theories \cite{Haouzi:2020yxy}, 
\begin{equation}
Z_{\text{defect}}^{(k)}(\{x_j\}_{j=1}^w)=\prod_{j=1}^w\sh^\delta(X_j)\prod_{i=1}^{N}\sh\left(X_j\pm a_{i}\right)\prod_{I=1}^{k}\frac{\sh\left(\pm\phi_{I}\pm X_j+\epsilon_{-}\right)}{\sh\left(\pm\phi_{I}\pm X_j-\epsilon_{+}\right)}.
\end{equation}
For Sp($N$) theories, we present the expression of $Z^{(k)}_{\rm defect}$ in Appendix \ref{a:Sp}. $w$ here denotes the number of defects inserted, i.e. number of D4${}^'$-branes. 

In the practical computation, the poles obtained from the Jeffrey–Kirwan (JK) prescription in the defect partition function consist of two families. One is the same as the set of poles in the partition function without defects, denoted as $C_{\rm JK}^{\rm pure}$, and the other comes from the poles in $Z^{(k)}_{\rm defect}$. One may then define an operator $Y(x)$ s.t. 
\begin{align}
    \left\langle[Y(x)]^{\pm1}\right\rangle=\frac{1}{Z^G_{\rm inst}}\sum_{k=0}^\infty\frac{\mathfrak{q}^k}{|W(\hat{G}_k)|}\oint_{C_{\rm JK}^{\rm pure}}\prod_{I=1}^k\frac{d\phi_I}{2\pi i}Z_{\text{vec}}^{(k)}\cdot\left[Z_{\text{defect}}^{(k)}(x)\right]^{\pm1},
\end{align}
and express the defect partition functions in terms of the expectation values of $Y^{\pm 1}$. 

When $w=1$, the qq-character gets simplified. In SU($N$) theories, it is possible to pick up at most one pole from $Z^{(k)}_{\rm defect}$, and we have a neat expression as \cite{Nekrasov:2015wsu,Bourgine:2015szm,Kimura:2015rgi,Bourgine:2016vsq,Bourgine:2017jsi}, 
\begin{align}
    \left\langle\chi^{{\rm SU}(N)}(x)\right\rangle:=\frac{Z_{\text{1 defect}}^{{\rm SU}(N)}(x)}{Z_{\text{inst}}^{{\rm SU}(N)}}=\langle Y(x)\rangle+\mathfrak{q}\left\langle\frac1{Y(x+2 \epsilon_+)}\right\rangle.\label{qq-Y}
\end{align}
By further expanding with an appropriate defect fugacity, we obtain the non-perturbative Schwinger-Dyson equations \cite{Nekrasov:2015wsu}. It resembles the fundamental character of SU(2), $\chi=y+y^{-1}$, where this SU(2) corresponds to the quiver structure of a pure gauge theory, and this is why the defect partition function is also called the (expectation value of) qq-character, $\chi^{{\rm SU}(N)}(x)$. It can also be understood as a double-quantization of the Seiberg-Witten curve. In SO and Sp gauge theories, it is difficult to have closed-form expressions for the qq-characters on the generic $\Omega$-background as \eqref{qq-Y} \cite{Haouzi:2020yxy}, but in the unrefined limit, one can still achieve it \cite{Nawata:2023wnk} by using the trick developed in the extension of the topological vertex formalism with the so-called O-vertex \cite{Hayashi:2020hhb,Nawata:2021dlk,Nawata:2023wnk,Kim:2024ufq,Kim:2025eal}. The qq-character with $w=1$ is related to the Wilson loops of antisymmetric representations as \eqref{qq-Wilson}. By expanding the qq-characters with larger $w$, one can access the Wilson loops of the tensor product of antisymmetric representations, and in SU($N$) theories, it gives all the Wilson loops in linear combinations.

\subsection{Examples}\label{s:w-example}

In this section, we provide detailed illustrations of these computational approaches through representative examples. We check the calculation for $\rm{SU}(N)$ with higher representations in Section~\ref{eq:sec3.3.1}. In Section~\ref{eq:sec3.3.2} we compute the Wilson loops for $\rm{SO}(5)$ and $\rm{Sp}(2)$ theories, and check their equivalence since they have the same Lie algebras. In Appendix~\ref{a:su4}, we present a similar comparison for $\rm{SU}(4)$ and $\rm{SO}(6)$ theories, and also the isomorphism between Sp(1) and SU(2) in Appendix \ref{a:su2}. In Section~\ref{eq:sec3.3.3}, we construct the Wilson loops for pure $G_2$ gauge theory.

\subsubsection{SU(N)}\label{eq:sec3.3.1}
Our primary focus is the 5d gauge theory with $A$-type gauge group $\mathrm{SU}(N)_{0}$ and Chern-Simons level $\kappa=0$.
The instanton partition function for $\mathrm{SU}(N)_{\kappa}$ theory is given by the sum over $k$-instanton contour integrals:
\begin{equation}
    Z_{\text{inst}}^{{\rm SU}(N)_{\kappa}}=\sum_{k=0}^{\infty}\frac{\mathfrak{q}^k}{k!}\oint\prod_{I=1}^{k}\frac{{\rm d}\phi_I}{2\pi i}\, e^{-\kappa \sum_{I=1}^{k}\phi_I}Z^{(k)}_{\textrm{vec}} ,\label{Z-preSU}
\end{equation}
where $Z_{\text{vec}}^{(k)}$ is defined in \eqref{eq:ZvecSUN}. The contributions from Wilson loop insertions in fundamental, anti-fundamental, and antisymmetric representations are respectively given by
\begin{align}
    Z_{\mathbf{F}}^{(k)}=&\chi_{\mathbf{F}}(A_i)\,e^{-\frac{1}{2}\sum_{I=1}^{k}\phi_I-\frac{k}{2}m_{\mathbf{F}}}\prod_{I=1}^{k}\sh(\phi_I+m_i),\\
    Z_{\overline{\mathbf{F}}}^{(k)}=&\chi_{\overline{\mathbf{F}}}(A_i)\,e^{\frac{1}{2}\sum_{I=1}^{k}\phi_I-\frac{k}{2}m_{\overline{\mathbf{F}}}}\prod_{I=1}^{k}\sh(-\phi_I+m_i),\\
    Z_{{\mathbf{\Lambda}^2}}^{(k)}=&\chi_{\mathbf{\Lambda}^2}(A_i)\,e^{-\frac{1}{2}(N-4)\sum_{I=1}^{k}\phi_I-\frac{k\,N}{2}m_{{\mathbf{\Lambda}}^2}}\prod_{I=1}^{k}\frac{\prod_{i=1}^N\sh(\phi_I+a_i+m_{\mathbf{\Lambda}^2})}{\sh(2\phi_I\pm\epsilon_++m_{\mathbf{\Lambda}^2})}\nonumber\\
    &\qquad\qquad\qquad\qquad\qquad\qquad\quad\cdot\prod_{I<J}^{k}\frac{\sh(\phi_I+\phi_J\pm\epsilon_-+m_{\mathbf{\Lambda}^2})}{\sh(\phi_I+\phi_J\pm\epsilon_++m_{\mathbf{\Lambda}^2})}.
\end{align}
These correspond to hypermultiplet insertions in respective representations with Chern-Simons levels $\kappa=1$, $-1$, $N-4$ respectively.

The $k$-instanton integral can be evaluated as a residue sum over Jeffrey-Kirwan (JK) poles. For pure SU($N$) theory without hypermultiplets, these poles are classified by Young diagrams $\mu_i$:
\begin{equation}
    \{\phi_I\}_{I=1}^k=\left\{\left.a_i+(m-1/2)\epsilon_1+(n-1/2)\epsilon_2\right|(m,n)\in \mu_i,i=1,\dots,N\right\},\label{Young-pole}
\end{equation}
Consequently, the instanton partition function admits a simple analytic expression given by 
\begin{equation}
    Z^{{\rm SU}(N)_{\kappa}}_{\rm inst}=\sum_{\{\mu_i\}_{i=1}^N}\left(\mathfrak{q}\left(q_1 q_2\right)^{N/2}\right)^{\sum_{\alpha=1}^N|\mu_\alpha|}\cdot\frac{e^{-\kappa \sum_{I=1}^{k}\phi_I}}{\prod_{I,J=1}^{N}\mathcal{N}_{\mu_I\mu_J}(Q_{IJ};q_1,q_2)},
\end{equation}
where $Q_{IJ}=A_I/A_J,\,A_I=e^{a_I}$ and $ \mathcal{N}_{\mu_1\mu_2}(Q;q_1,q_2)$ represents the Nekrasov factors, 
 \begin{align}
    \mathcal{N}_{\mu_1\mu_2}(Q;q_1,q_2):=\prod_{(m,n)\in\mu_1}\left(1-Q q_1^{-\mu_{1,m}+n} q_2^{\mu^t_{2,n}-m+1}\right)\cdot\prod_{(m,n)\in\mu_2}\left(1-Qq_1^{\mu_{2,m}-n+1}q_2^{-\mu^t_{1,n}+m}\right).\,
\end{align}
By adding $N_{\mathbf{F}}$ fundamental and $N_{\overline{\mathbf{F}}}$ hypermultiplets, the pole structure of the contour integral doesn't change, but there could be extra factors. We observe that, with the maximal insertions $N_{\mathbf{F}}=N_{\overline{\mathbf{F}}}=N$ and Chern-Simons level $0$, the partition function has the extra factor:
\begin{align}
    Z_{\text{extra}}=\mathrm{PE}\left[-\frac{\sqrt{q_1q_2}}{(1-q_1)(1-q_2)}\left(e^{\epsilon_++\sum_i m_{i}}+e^{-\epsilon_++\sum_i \bar{m}_{i}}\right)\right],
\end{align}
from which we can derive that the VEVs for Wilson loops $\langle W_{[r_1,0,\cdots,0,r_2]}\rangle$ in the representation $[r_1,0,\cdots,0,r_2]=\mathbf{F}{}^{\otimes r_1}\otimes\overline{\mathbf{F}}{}^{\otimes r_2},\,r_1,r_2\leq N$ via
\begin{align}
    \langle W_{[r_1,0,\cdots,0,r_2]}\rangle =\frac{1}{Z^{{\rm SU}(N)_{\kappa}}_{\rm inst}}\sum_{\{\mu_i\}_{i=1}^N}W_{[r_1,0,\cdots,0,r_2],\mu}\frac{e^{\kappa \sum_{I=1}^{k}\phi_I}\,\left(\mathfrak{q}\left(q_1 q_2\right)^{N/2}\right)^{\sum_{\alpha=1}^N|\mu_\alpha|}}{\prod_{I,J=1}^{N}\mathcal{N}_{\mu_I\mu_J}(Q_{IJ};q_1,q_2)},
\end{align}
for $\kappa=\frac{1}{2}(r_1-r_2)$. The insertion 
\begin{align}
    W_{[r_1,0,\cdots,0,r_2],\mu}=\mathrm{Ch}_{\mathbf{F},\mu}^{r_1}\mathrm{Ch}_{\overline{\mathbf{F}},\mu}^{r_2}+\mathcal{I}^{N-1}\left(\delta_{r_1,N}/\sqrt{q_1q_2}+\delta_{r_2,N}\sqrt{q_1q_2}\right)\mathfrak{q},
\end{align}
where $\mathrm{Ch}_{\mathbf{F},\mu}$ is defined as the equivariant Chern character in the fundamental representation evaluated at JK-poles:
\begin{align}
    \Ch_{\mathbf{F}}(A_I,q_1,q_2)=\sum_{I=1}^N A_I-\frac{(1-q_1)(1-q_2)}{(q_1 q_2)^{\frac{1}{2}}}\sum_{I=1}^{N}\left(A_I\cdot\sum_{(m,n)\in \mu_I}q_1^{n-\frac{1}{2}}q_2^{-\frac{1}{2}+m}\right).
\end{align}
and $\mathrm{Ch}_{\overline{\mathbf{F}},\mu}$ is defined as
\begin{align}
    \Ch_{\overline{\mathbf{F}}}(A_I,q_1,q_2)=\Ch_{\mathbf{F}}(A_I^{-1},q_1^{-1},q_2^{-1}).
\end{align}
By explicitly calculating the VEVs of Wilson loops and utilizing the BPS expansion, we observe that the free energy of Wilson loops at one instanton has a universal expression
\begin{align}\label{eq:Ffund}
    &\mathcal{F}_{[r_1,0,\cdots,0,r_2]}^{(1)}=\nonumber\\
    &\qquad\mathcal{I}^{\,r_1+r_2-1}(-1)^{\frac{1}{2}(N+r_1+r_2-1)}\left((-1)^{N}v^{2-2N}\delta+\sum_{k=-\min(r_1,r_2)}^{\infty}\chi_{(r_1+k,0,\cdots,0,r_2+k)}v^{2k}\right),
\end{align}
where $\delta=1$ if $r_1=N, r_2=0$ or $r_1=0, r_2=N$ and $\delta=0$ for all other cases. The expression \eqref{eq:Ffund} resembles the Hilbert series expansion for the one-instanton partition function of pure gauge theories. In particular, if the gauge fugacities are 0, it has the structure:
\begin{align}
    \mathcal{F}_{\mathbf{r}}^{(1)}=\mathcal{I}^{\,\sum_ir_i-1}\frac{v^{2h^{\vee}_G-2}f(v)}{(1-v^{2})^{2h^{\vee}_G-2}},
\end{align}
such that $f(v)=f(v^{-1})$.

We also investigated Wilson loops in representations that contain the antisymmetric representation. In this case, we need to consider the partition function of $\mathrm{SU}(N)_{\kappa}+{\mathbf{\Lambda}}^2+N_{\mathbf{F}}\,\mathbf{F}+N_{\overline{\mathbf{F}}}\,\overline{\mathbf{F}}$ theory with Chern-Simons level $\kappa=\frac{1}{2}(N-4+N_{\mathbf{F}}-N_{\overline{\mathbf{F}}})$. For $N\geq 3$, we observe that if $N_{\mathbf{F}}=4$ and $N_{\overline{\mathbf{F}}}=N-1$, the extra factor is
\begin{align}
    Z_{\text{extra}}=\mathrm{PE}\left[-\frac{\sqrt{q_1q_2}}{(1-q_1)(1-q_2)}\left(e^{\epsilon_++\sum_i m_{i}}\right)\right],
\end{align}
which does not depend on $N$ and the Coulomb parameters. 
The one-instanton results for the Wilson loops by turning off all gauge fugacities are presented in Appendix \ref{a:su(n)-k}.

\subsubsection{SO(5) and Sp(2)}\label{eq:sec3.3.2}
In this section, we compute the Wilson loop VEVs for 5d $\mathfrak{so}(5)$ and $\mathfrak{sp}(2)$ theories \footnote{To be more precise, we shall consider Spin(5) gauge theory instead of SO(5) to include Wilson loops in the spinor representation. However, we will not emphasize this point too much, since the computation of Nekrasov partition function only relies on the Lie-algebraic structure. }. They have the same Lie algebra, so their calculations are expected to be identical under maps of representations.
In the case of $\mathfrak{so}(5)\cong \mathfrak{sp}(2)$, it is tricky to construct Wilson loops of general representations. It is not straightforward to compute Wilson loops of mixed tensor product of the fundamental representations ${\bf 4}$ and ${\bf 5}$, since the qq-characters only give representations of antisymmetric products of the smallest fundamental representations (in $\mathfrak{so}(5)$ the vector representation ${\bf 5}$, and in $\mathfrak{sp}(2)$ the ${\bf 4}$-dim representation). In this section, we take the Chern-character approach and start from the Sp(2) theory side to compute an arbitrary Wilson loop in these isomorphic theories.

The ADHM quantum mechanics for Sp-theories has O$_{\pm}(k)$ sectors (see Appendix \ref{a:Sp} for the integrands in Sp-theories). Denote $\Ch^\pm_{\bf 4}$ as the equivariant Chern characters for the O$_{\pm}(k)$ sectors.
In the fundamental representation ${\bf 4}$, their expressions have been given in \cite[eq. (D.15)]{Kim:2012gu} \footnote{For the ${\rm O}_-(2n)$ sector, the expression $\Ch^{-}_{\mathbf{4}}$ we present here is different from \cite[eq. (7.32)]{Gaiotto:2015una}. The result \eqref{eq:CHsp_minus} is derived from the large mass expansion of the fundamental hypermultiplet contribution in \cite[eq. (A.19)-(A.21)]{Gaiotto:2015una} . }. For the ${\rm O}_+(k)$ sector with $k=2n+\chi$,
\begin{align}
    &\Ch^{+}_{\mathbf{4}}(A_i,e^{-\phi_j},q_1,q_2, \chi)=\sum_{I=1}^2\left(A_I+\frac{1}{A_I}\right)-\frac{(1-q_1)(1-q_2)}{(q_1q_2)^{\frac{1}{2}}}\left(\sum_{I=1}^{n} \left(e^{\phi_I}+e^{-\phi_I}\right) +\chi\right),
\end{align}
and for ${\rm O}_-(k)$ with $k=2n+1$,
\begin{align}
    &\Ch^{-}_{\mathbf{4}}(A_i,e^{-\phi_j},q_1,q_2, e^{i\pi})=\sum_{I=1}^2\left(A_I+\frac{1}{A_I}\right)-\frac{(1-q_1)(1-q_2)}{(q_1q_2)^{\frac{1}{2}}}\left(\sum_{I=1}^{n} \left(e^{\phi_I}+e^{-\phi_I}\right) +e^{i\pi}\right),
\end{align}
for ${\rm O}_-(k)$ with $k=2n$,
\begin{align}\label{eq:CHsp_minus}
    &\Ch^{-}_{\mathbf{4}}(A_i,e^{-\phi_j},q_1,q_2, e^{i\pi})=\sum_{I=1}^2\left(A_I+\frac{1}{A_I}\right)-\frac{(1-q_1)(1-q_2)}{(q_1q_2)^{\frac{1}{2}}}\left(\sum_{I=1}^{n-1} \left(e^{\phi_I}+e^{-\phi_I}\right) +1+e^{i\pi}\right).
\end{align}
Using the fact that $\bigwedge^2{\bf 4}={\bf 5}\oplus {\bf 1}$, one can construct the Chern character for ${\bf 5}$ representation in Sp(2) theory as
\begin{align}
    \Ch^\pm_{\bf 5}&(A_i,e^{-\phi_j},q_1,q_2,\chi)=\cr
    &\frac{1}{2}\left(\Ch^\pm_{\bf 4}(A_i,e^{-\phi_j},q_1,q_2,\chi)^2-\Ch^\pm_{\bf 4}(A_i^2,e^{-2\phi_j},q^2_1,q^2_2,(\chi)^2)\right)
    -1,
\end{align}
Note that for the ${\rm O}_-(k)$ sector, we always use the notation $\chi=e^{i\pi}$. For instance, at the level of $k=2$, $\Ch^-_{\bf 5}$ is calculated as
\begin{align}
    \left.\Ch^-_{\bf 5}(A_i,q_1,q_2,\chi=0)\right|_{k=2}
    &=\frac{1}{2}\left(\Ch^-_{\bf 4}(A_i,q_1,q_2,e^{i\pi})^2-\Ch^-_{\bf 4}(A_i^2,q^2_1,q^2_2,e^{2i\pi})\right)-1\cr
    &=A_1A_2+\frac{A_1}{A_2}+\frac{A_2}{A_1}+\frac{1}{A_1A_2}+1+\frac{(1-q^2_1)(1-q^2_2)}{q_1q_2}.
\end{align}
We checked the agreement between the Wilson loop with ${\bf 5}$-dim representation in $\mathfrak{sp}(2)$ and that of the vector representation in $\mathfrak{so}(5)$ up to 3 instantons. 
Using the combinations of the above Chern characters, we can compute the Wilson loops of tensor products of ${\bf 4}$ and ${\bf 5}$. This enabled us to check the blowup equations shown in Section \ref{s:blowup-so5}.

\subsubsection{$G_2$}\label{eq:sec3.3.3}
In this section, we compute the Wilson loops in 5d pure $G_2$ gauge theory, which will be used in the establishment of blowup equations in Section \ref{s:blowup-G2}. We apply the ADHM-like construction proposed in \cite{Kim:2016foj,Kim:2018gjo} to study the codimension-4 defect partition function of 5d pure $G_2$ gauge theory and extract the Wilson loops from them. Although the ADHM construction is only possible for classical gauge groups, the key idea \cite{Kim:2018gjo} is to Higgs an SO(7) gauge theory with a spinor hypermultiplet ${\bf 8}$ to obtain a pure $G_2$ gauge theory (or even adding fundamental matters ${\bf 7}$ of $G_2$). In the realization of SO(7), a further trick is to use the ADHM construction of SU(4) gauge theory (plus a vector multiplet in representation ${\bf 6}$, as the adjoint of SO(7), ${\bf 21}$, is decomposed into the adjoint and the antisymmetric of SU(4), ${\bf 15}\oplus{\bf 6}$). Then, after the Higgsing to $G_2$, the JK prescription reduces to that of SU(3), and all the residues contributing to the $G_2$ partition function are labeled by three Young diagrams $\vec{Y}$. More precisely, the $k$-instanton partition function is given by, 
\begin{eqnarray}\label{G2-part}
  Z_{k}^{G_2}&=&\frac{1}{k!}\oint_{C_{\rm JK}}\prod_{I=1}^k\frac{{\rm d}{\phi}_I}{2\pi i}\cdot
  \frac{\prod_{I\neq J}\sh({\phi_{IJ}})\cdot
  \prod_{I,J}\sh({2\epsilon_+-{\phi}_{IJ}})}
  {\prod_{I=1}^k\prod_{i=1}^3 {\sh}({\epsilon_+\pm({\phi}_I-v_i)})
  \cdot\prod_{I,J}\sh({\epsilon_{1,2}+{\phi}_{IJ}})}\nonumber\\
  &&\times\frac{\prod_{I\leq J}\left(\sh({{\phi}_I+{\phi}_J})
  \cdot \sh({{\phi}_I+{\phi}_J-2\epsilon_+})\right)}
  {\prod_I\left(\left(\prod_{i=1}^3 \sh({\epsilon_+-{\phi}_I-v_i})\right)
  \cdot \sh({\epsilon_+-{\phi}_I})\right)
  \cdot\prod_{I<J}\sh({\epsilon_{1,2}-{\phi}_I-{\phi}_J})}\cr
  &&=(-1)^k\sum_{\vec{Y};|\vec{Y}|=k}\prod_{i=1}^3\prod_{s\in Y_i}
  \frac{1}
  {\prod_{j=1}^3\left(\sh({E_{ij}(s)})\cdot
  \sh({E_{ij}(s)-2\epsilon_+})\right)}\nonumber\\
  \hspace*{-1cm}&&\times\frac{\sh(2{u}(s))\cdot \sh(2\epsilon_+-2{u}(s))}
  {\prod_{j=1}^3 \sh({\epsilon_+-{\phi}(s)-v_j})
  \cdot \sh({\epsilon_+-{\phi}(s)})}\nonumber\\
  \hspace*{-1cm}&&\times\prod_{i\leq j}^3\prod_{s_{i,j}\in Y_{i,j};s_i<s_j}
  \frac{\sh({{\phi}(s_i)+{\phi}(s_j)})\cdot
  \sh({{\phi}(s_i)+{\phi}(s_j)-2\epsilon_+})}
  {\sh({\epsilon_{1,2}-{\phi}(s_i)-{\phi}(s_j)})},
\end{eqnarray}
where the Coulomb branch parameters in SU(3), $\vec{v}$, are given in terms of the $G_2$ Coulomb branch parameters $a_1$ and $a_2$ as 
\begin{equation}
    \vec{v}=(-a_1-a_2,-a_1,a_2+2a_1).
\end{equation}
$\phi(s)$ is defined in the same way as \eqref{Young-pole},  
\be
{\phi}(s)=v_i-\epsilon_+-(n-1)\epsilon_1-(m-1)\epsilon_2,\quad s=(m,n)\in Y_i,
\ee
and 
\be
E_{ij}(s)= v_i-v_j-\epsilon_1 h_i(s)+\epsilon_2(\ell_j(s)+1),
\ee
Here and below, $s_i\!<\!s_j$ means ($i<j$) or ($i\!=\!j$ and $m_i\!<\!m_j$) or
($i\!=\!j$ and $m_i\!=\!m_j$ and $n_i\!<\!n_j$).
$h_i(s)$, usually called arm, denotes the horizontal distance from $s$ to the right end of the diagram $Y_i$. $\ell_j(s)$, called leg, stands for the
vertical distance from $s$ to the bottom of the diagram $Y_j$.

There are again two approaches to the computation of the Wilson loops. One is to insert the Chern characters of the fundamental representation. Since in the decomposition $G_2\to{\rm SU}(3)$, the fundamental representation of $G_2$ decomposes as ${\bf 7}\to {\bf 3}\oplus\bar{\bf 3}\oplus {\bf 1}$, one can insert the following Chern character 
\begin{equation}
    \Ch^{G_2}_{\bf 7}(A_I,q_1,q_2)=\Ch^{\rm SU(3)}_{\bf 3}(V_I,q_1,q_2)+\Ch^{\rm SU(3)}_{\bar{\bf 3}}(V_I,q_1,q_2)+1,
\end{equation}
to generate Wilson loops of tensor representation of ${\bf 7}$ in $G_2$ theory. One can also mimic the way people constructed the Chern character for antisymmetric representation in SU($N$) theories (refer to e.g. \cite[A.15]{Gaiotto:2015una}) to include the adjoint Chern character in $G_2$, 
\begin{equation}
    \Ch_{\bf 14}^{G_2}(A_I,q_1,q_2)=\frac{1}{2}\left(\Ch_{\bf 7}^{G_2}(A_I,q_1,q_2)^2-\Ch_{\bf 7}^{G_2}(A^2_I,q_1^2,q_2^2)\right)-\Ch_{\bf 7}^{G_2}(A_I,q_1,q_2).
\end{equation}
We note that it is also necessary to include the extra factors as proposed in \eqref{extra-factor} when we compute Wilson loops of large tensor products of fundamental and adjoint representations. 

The second approach is via the qq-character computation. Recall that in a pure SU($N$) gauge theory, we introduce the contribution \eqref{defect-SU} to compute the codimension-4 defect partition function, and in $G_2$ theory, due to the same reason that ${\bf 7}\to {\bf 3}\oplus\bar{\bf 3}\oplus {\bf 1}$, we shall realize each defect contribution in $G_2$ by two copies of SU(3) defects as 
\begin{equation}\label{defect1}
Z^{(k)}_{G_2\ {\rm defect}}=\prod_{i=1}^{3}\sh(x-v_i)\prod_{I=1}^k\frac{\sh(\pm(\phi_I-x)+\epsilon_-)}{\sh(\pm(\phi_I-x)-\epsilon_+)}.
\end{equation}
The full defect partition (qq-character) is then given by the integral, 
\begin{eqnarray}\label{G2-integral}
  Z_{G_2\ {\rm defect}}&=&\sum_{k=0}^\infty\frac{1}{k!}\oint_{C_{\rm JK}}\prod_{I=1}^k\frac{d{\phi}_I}{2\pi i}\cdot
  \frac{\prod_{I\neq J}\sh({\phi_{IJ}})\cdot
  \prod_{I,J}\sh({2\epsilon_+-{\phi}_{IJ}})}
  {\prod_{I=1}^k\prod_{i=1}^3 {\sh}({\epsilon_+\pm({u}_I-v_i)})
  \cdot\prod_{I,J}\sh({\epsilon_{1,2}+{\phi}_{IJ}})}\nonumber\\
  &&\times\frac{\prod_{I\leq J}\left(\sh({{\phi}_I+{\phi}_J})
  \cdot \sh({{\phi}_I+{\phi}_J-2\epsilon_+})\right)}
  {\prod_I\left(\left(\prod_{i=1}^3 \sh({\epsilon_+-{\phi}_I-v_i})\right)
  \cdot \sh({\epsilon_+-{\phi}_I})\right)
  \cdot\prod_{I<J}\sh({\epsilon_{1,2}-{\phi}_I-{\phi}_J})}\nonumber\\
  &&
  \times \prod_{i=1}^{3}{\sh(\pm x-v_i)}\prod_{I=1}^k\frac{\sh(\pm(\phi_I\pm x)+\epsilon_-)}{\sh(\pm(\phi_I\pm x)-\epsilon_+)}.
\end{eqnarray}
The residue sum can be labeled by five Young diagrams growing around the following five parameters, 
\be
\vec{v}=(-a_1-a_2,-a_1,a_2+2a_1,-x+2\epsilon_+,x+2\epsilon_+),
\ee
and the defect partition function is evaluated to 
\begin{eqnarray}\label{G2-residue}
  Z_{G_2\ {\rm defect}}\!&\!=\!&\!\sum_{k=0}^\infty(-1)^k\sum_{\vec{Y};|\vec{Y}|=k}\prod_{i=1}^5\prod_{s\in Y_i}
  \frac{1}
  {\prod_{j=1}^5\left(\sh({E_{ij}(s)})\cdot
  \sh({E_{ij}(s)-2\epsilon_+})\right)}\nonumber\\
  \hspace*{-1cm}&&\cdot\frac{\sh(2{\phi}(s))\cdot \sh(2\epsilon_+-2{\phi}(s))}
  {\prod_{j=1}^3 \sh({\epsilon_+-{\phi}(s)-v_j})
  \cdot \sh({\epsilon_+-{\phi}(s)})}\nonumber\\
  \hspace*{-1cm}&&\cdot\prod_{i\leq j}^5\prod_{s_{i,j}\in Y_{i,j};s_i<s_j}
  \frac{\sh({{\phi}(s_i)+{\phi}(s_j)})\cdot
  \sh({{\phi}(s_i)+{\phi}(s_j)-2\epsilon_+})}
  {\sh({\epsilon_{1,2}-{\phi}(s_i)-{\phi}(s_j)})}\nonumber\\
  \hspace*{-1cm}&&\times \prod_{i=1}^{3}{\sh(\pm x-v_i)}\prod_{i=1}^5\prod_{s\in Y_i} \frac{\sh(3\epsilon_+-\phi(s)\pm x)\sh(\epsilon_-\pm \phi(s)\pm x)}{\sh(-\phi(s)\pm x -\epsilon_+)}.
\end{eqnarray}
The Wilson loops of antisymmetric products of fundamental representations can then be obtained from the above defect partition function (i.e. qq-character).

\section{Blowup equations}\label{s:blowup}

The blowup equations are a set of identities satisfied by the partition functions with/without Wilson loops. The formulation was first proposed in a series of works by Nakajima and Yoshioka \cite{Nakajima:2003pg,Nakajima:2005fg,Nakajima:2009qjc} in the study of instanton partition functions of 4d ${\cal N}=2$ theories, and then generalized to other theories especially in 5d and 6d in many follow-up works \cite{Gu:2017ccq,Huang:2017mis,Gu:2018gmy,Bershtein:2018zcz,Gu:2019dan,Kim:2019uqw,Gu:2019pqj,Shchechkin:2020ryb,Gu:2020fem,Kim:2021gyj,Wang:2023zcb,Jeong:2020uxz,Nekrasov:2020qcq,Kim:2020hhh,Kim:2023glm}. In this section, we first give the prescription to explicitly write down the blowup equations involving Wilson loops based on the generalization of \cite{Wang:2023zcb,Kim:2021gyj}, and then we check them explicitly with the results obtained in previous sections.

Consider a 5d pure gauge theory without hypermultiplets. The blowup equation relates the partition function on the blown-up background $S^1\times \widehat{\mathbb{C}}^2$ and the partition functions of Wilson loops on the background $S^1\times \mathbb{C}^2$. 
Schematically, the partition function of $S^1\times \widehat{\mathbb{C}}^2$ is given by 
\begin{equation}
    \hat{Z}(b,{\bf r}_{\rm N},{\bf r}_{\rm S})=\sum_{\vec{n}\in{\bf \Lambda}}f_{b,\vec{\omega}^\vee}(\vec{n})Z^{({\rm N})}_{W_{{\bf r}_{\rm N}}}(\vec{n}+\vec{\omega}^\vee,b)Z^{({\rm S})}_{W_{{\bf r}_{\rm S}}}(\vec{n}+\vec{\omega}^\vee,b).\label{blowup-eq}
\end{equation}
Here ${\bf \Lambda}$ refers to the coroot lattice. $\omega^\vee$ can be set to zero or one of the fundamental coweights, and the corresponding coefficient $f_{b,\vec{\omega}^\vee}$ is given by 
\begin{align}
    f_{b,\vec{\omega}^\vee}(\vec{n})=\lt(\mathfrak{q}(q_1q_2)^{\frac{b}{2}}\rt)^{\frac{(\vec{n}+\vec{\omega}^\vee)\cdot (\vec{n}+\vec{\omega}^\vee)-\vec{\omega}^\vee\cdot \vec{\omega}^\vee}{2}}e^{-\frac{b-h^\vee}{2}\vec{a}\cdot (\vec{n}+\vec{\omega}^\vee)}\prod_{\alpha\in\Delta}{\cal L}_{(\vec{n}+\vec{\omega}^\vee)\cdot \vec{\alpha}}(\vec{a}\cdot \vec{\alpha},\epsilon_1,\epsilon_2)^{-1},\label{f-factor}
\end{align}
where 
\begin{equation}
    {\cal L}_k(x,\epsilon_1,\epsilon_2)=\lt\{
    \begin{array}{cc}
        \prod_{m,n=0}^{m+n\leq k-2}\lt(1-q_1^{m+1}q_2^{n+1}e^{-x}\rt) & {\rm for}\ k\geq 2,\\
        \prod_{m,n=0}^{m+n\leq -k-1}\lt(1-q_1^{-m}q_2^{-n}e^{-x}\rt) & {\rm for}\ k\leq -1,\\
        1 & {\rm for}\ k=0,1.
    \end{array}
    \rt.
\end{equation}
$b$ is a free parameter that can be chosen as $b=h^\vee\mod 2$, and when we choose $\omega^\vee=0$, the above coefficient $f_{b,\vec{\omega}^\vee}$ is exactly the same as that given in \cite{Kim:2019uqw}. $Z^{{\rm N}/{\rm S}}$ are respectively given by 
\begin{align}
    Z^{({\rm N})}_{W_{\bf r}}(\vec{k},b)=Z_{W_{\bf r}}(\vec{a}+\vec{k}\epsilon_1,\epsilon_1,\epsilon_2-\epsilon_1;\mathfrak{q} q_1^{\frac{b}{2}}),\\
    Z^{({\rm S})}_{W_{\bf r}}(\vec{k},b)=Z_{W_{\bf r}}(\vec{a}+\vec{k}\epsilon_2,\epsilon_1-\epsilon_2,\epsilon_2;\mathfrak{q} q_2^{\frac{b}{2}}),
\end{align}
where $Z_{W_{\bf r}}$ stands for the unnormalized Wilson loop (i.e. the VEV of the Wilson loop $\langle W_{\bf r} \rangle$ multiplied by the instanton partition function). If the representation $\mathbf{r}=\mathbf{1}$ is trivial, $Z_{W_{\bf r}}$ stands for the instanton partition function. 

We note that the factors $f_{b,\vec{\omega}^\vee}$ and ${\cal L}_k$ can be derived from the perturbative part of the partition function. More precisely, the full partition function $\mathcal{Z}_{W_{\bf r}}$ is factorized into a classical contribution $Z_{\rm class}$, the one-loop contribution $Z_{\rm 1-loop}$, and the remaining part $Z_{W_{\bf r}}$, 
\begin{equation}
    \mathcal{Z}_{W_{\bf r}}=Z_{\rm class}Z_{\rm 1-loop}Z_{W_{\bf r}}.
\end{equation}
where the expressions for $Z_{\rm class}$ and $Z_{\rm 1-loop}$ are the contributions for the partition function, which can be found e.g. in \cite[(2.22)-(2.24)]{Kim:2019uqw}. 
The blowup equation for the full partition function is given by
\begin{equation}
    \hat{{\cal Z}}(b,{\bf r}_{\rm N},{\bf r}_{\rm S})=\sum_{\vec{n}\in{\bf \Lambda}}{\cal Z}^{({\rm N})}_{W_{{\bf r}_{\rm N}}}(\vec{n}+\vec{\omega}^\vee,b){\cal Z}^{({\rm S})}_{W_{{\bf r}_{\rm S}}}(\vec{n}+\vec{\omega}^\vee,b),\label{bear-blowup}
\end{equation}
and dividing $Z_{\rm class}Z_{\rm 1-loop}$ to the r.h.s. gives $f_{b,\vec{\omega}^\vee}$. The Wilson-loop partition function $Z_{W_{{\bf r}}}$ then starts from the character of the corresponding representation at its zero-instanton level, 
\begin{equation}
    Z_{W_{{\bf r}}}=\chi_{{\bf r}}+{\cal O}(\mathfrak{q}),\qquad \chi_{{\bf r}}=\sum_{\omega\in \mathbf{r}}e^{-\omega\cdot a}\label{zero-inst}
\end{equation}

Schematically, the blowup equation can be derived as follows. Consider a partition function (with/without defect) on $\mathbb{C}^2$. Its computation localizes to the contribution at the origin. One can blow up one point on $\mathbb{C}^2$ and perform the localization computation again to obtain a product of the instanton contributions from the north and south poles of the blowup $\mathbb{P}^1$ with summation over all allowed non-trivial flux on $\mathbb{P}^1$. This gives the blowup equation \eqref{bear-blowup} and \eqref{blowup-eq}.

The reason that the equation \eqref{blowup-eq} or \eqref{bear-blowup} is named as blowup equation is that the partition function $\hat{Z}$ on the blown-up space is related to the partition function or VEVs of observables on the original spacetime $S^1\times\mathbb{C}^2$.  More specifically, for the cases we are interested in, $\hat{Z}$ is identical to a linear summation over partition functions with the insertion of Wilson loops on $S^1\times\mathbb{C}^2$:
\begin{equation}
    \hat{Z}(b,{\bf r}_{\rm N},{\bf r}_{\rm S})=\sum_i c_i^{(b)}(\mathfrak{q};\epsilon_1,\epsilon_2)Z_{W_{\mathbf{r}_i}}(\vec{a},\epsilon_1,\epsilon_2;\mathfrak{q}),\label{Zhat-expansion}
\end{equation}
where the coefficients $c^{(b)}(\mathfrak{q};\epsilon_1,\epsilon_2)$ are polynomials of $\mathfrak{q}$ \footnote{As has been studied in the elliptic blowup equations for 6d SCFTs or their 5d KK theories \cite{Gu:2018gmy,Gu:2019dan,Gu:2019pqj,Gu:2020fem,Kim:2021gyj}, the $c^{(b)}$ can be theta functions, e.g. infinite summation over $\frak{q}$.} and Laurent polynomials of $q_1,q_2$. The structure presented in \eqref{Zhat-expansion} was first explicitly observed in \cite{Kim:2021gyj} and later generalized to the present form in \cite{Wang:2023zcb} from the connection to topological string theory \footnote{See \cite{tian:2025yrj} for the connection to one-form symmetry from topological string B-models.} and explained in \cite{Bonelli:2025juz}. In the last part of this subsection, we will derive that the partition function on the left-hand side of \eqref{Zhat-expansion} has a discrete charge under the one-form symmetry transformation for the gauge field. This gives further constraints for the Wilson loop expansion on the right-hand side of \eqref{Zhat-expansion} that they must have the same charge under one-form symmetry actions.

For various given input $\{b,\omega^{\vee},\mathbf{r}_{\rm N},\mathbf{r}_{\rm S}\}$, if the coefficients $c^{(b)}(\mathfrak{q})$ are determined, the blowup equations provide a set of functional equations for the partition functions and VEVs of Wilson loops on $S^1\times \mathbb{C}^2$, which can be used to bootstrap their instanton contributions. 
To determine $c^{(b)}(\mathfrak{q})$, we need to address that the representations in the summation in \eqref{Zhat-expansion} has a top-weight, hence the summation is finite.
By substituting \eqref{zero-inst} into the r.h.s. of the blowup equation, we see that the representation of the top-weight Wilson loop in \eqref{Zhat-expansion} is determined by 
\begin{equation}
\mathbf{r}_{\mathrm{top}}=\lt\{
\begin{matrix}
    {\bf r}_\omega^{\otimes \frac{b-h^\vee}{2}}\otimes {\bf r}_{\rm N}\otimes {\bf r}_{\rm S} & b\geq h^\vee\\
    \overline{\bf r}_\omega^{\otimes \frac{-b-h^\vee}{2}}\otimes {\bf r}_{\rm N}\otimes {\bf r}_{\rm S} & b\leq -h^\vee
\end{matrix}
\rt.,\label{blowup-leading}
\end{equation}
where ${\bf r}_\omega$ and $\overline{\bf r}_\omega$ stand respectively the fundamental representation of weight $\omega$ and its complex conjugate. When $|b|<h^\vee$, the weight of the top Wilson loop gets reduced from ${\bf r}_{\rm N}\otimes {\bf r}_{\rm S}$ by an integer multiple ($|b-h^\vee|$) of $\omega$, recovering the expression presented in \cite{Kim:2021gyj}, and if the reduction cannot give a non-negative weight, the blowup equation will be vanishing, i.e. $\hat{Z}(b,{\bf r}_{\rm N},{\bf r}_{\rm S})=0$. This can be inferred from the factor $e^{-\frac{b-h^\vee}{2}\vec{a}\cdot \vec{n}}$ in the coefficient \eqref{f-factor}, and easily proved from the zero-instanton computation. 

The remaining terms in the expansion \eqref{Zhat-expansion} have {\it lower representations} \footnote{For a representation $\mathbf{r}$, let its highest weight (in the simple roots basis) be $[k_1,\cdots,k_{\mathrm{rank}(G)}]$. We call $\mathbf{r}^{\prime}$ a {\it lower representation} of $\mathbf{r}$ if its highest weight $[k_1^{\prime},\cdots,k_{\mathrm{rank}(G)}^{\prime}]$ satisfies $k_i^{\prime}\leq k_i$ for all $i$, and $k_i^{\prime}< k_i$ for at least one $i$.}, but should share the same one-form symmetry charge as the top-weight representation. Suppose we have recursively obtained the partition function and all the VEVs of Wilson loops with lower representations than the top representation $\mathbf{r}_{\rm top}$, in the limit  $t_i=\alpha_i\cdot a \rightarrow \infty$, the whole equation only has singular terms. If we know the singular structure for $Z_{W_{\mathbf{r}_{\mathrm{top}}}}$, since $c_i^{(b)}$ are independent of $a_i$, they shall be completely determined in this limit. By properly conjecturing the singular structure of $Z_{W_{\mathbf{r}_{\mathrm{top}}}}$ from its BPS sector $\mathcal{F}_{{\mathbf{r}_{\mathrm{top}}}}$, the instanton contribution for the VEVs of the top weight shall be bootstrapped from the blowup equations.

\paragraph{One-form symmetry charges on the blowup equations}
When $b\neq h^\vee$, the top-weight has different one-form symmetry charge from the r.h.s., i.e. ${\bf r}_{\rm N}\otimes {\bf r}_{\rm S}$. This is due to the fact that the l.h.s. of the blowup equation computes the defect partition function $\hat{Z}=\langle ({\cal O}_{\mathbb{P}^1})^d\rangle$ \cite{Kim:2019uqw,Nakajima:2003pg,Nakajima:2005fg} with $d=\frac{b-h^\vee}{2}$ and ${\cal O}_{\mathbb{P}^1}$ given by \cite{Baulieu:1997nj,Losev:1995cr}, 
\begin{align}
    {\cal O}_{\mathbb{P}^1}=&\exp\left[\int_{S^1\times M_4}\left(\overline{\omega}\wedge {\rm tr}(A\wedge {\rm d}A+\frac{2}{3}A\wedge A\wedge A)\right.\right.\cr
    &\left.\left.+\overline{\omega} \wedge \left(\phi {\rm tr}F+\frac{1}{2}\psi\wedge \psi\right)\wedge {\rm d}t+H{\rm tr}(F\wedge F)\wedge {\rm d}t\right)\right],\label{eq:Op1}
\end{align}
where $\overline{\omega}$ is a K\"ahler two-form on a $\mathbb{P}^1$ in $M_4$. One can indeed check that $({\cal O}_{\mathbb{P}^1})^d$ carries the same amount of one-form symmetry charge as the leading Wilson loop shown in \eqref{blowup-leading}. To prove this statement, we note that the corresponding exceptional divisor satisfies the condition \cite{Nakajima:2005fg}
\begin{equation}
    \langle c_1(F),[C]\rangle=-q.\label{pairing}
\end{equation}
$c_1(F)$ denotes the first Chern class of the gauge field $A$, and the pairing $\langle \bullet,\bullet\rangle$ is defined on $M_4$. $q$ corresponds exactly to the charge of the Wilson loop, whose representation is given by the fundamental representation that characterizes the coweight lattice in the blowup equation. Under the one-form symmetry (or the center symmetry $\mathbb{Z}_N$ of the gauge group), $A$ transforms as 
\begin{equation}
    A\to h_kAh^{-1}_k+h_k{\rm d}h^{-1}_k,
\end{equation}
with the constraint $h_k(x^0=R)=e^{\frac{2\pi k}{N}}h_k(x^0=0)$ ($k=0,1,\dots,N-1$) when we go around the circle $S^1$ with radius $R$, and the choice $k$ corresponds to the choice of coweight lattice appearing in the blowup equation. The first line of \eqref{eq:Op1} looks almost identical to the Chern-Simons action, and it is known that we obtain an additional (multiplicative) contribution after the gauge transformation as\footnote{See e.g. \cite[Exercise 2.6.2]{Dunne:1998qy}. We ignored a winding number term, which vanishes after the integration over $\overline{\omega}$. We also note that only the abelian part contributes to the pairing between the K\"ahler form and the first Chern class, $\int_{M^4}\overline{\omega}\wedge {\rm tr}F=\int_{M^4}\overline{\omega}\wedge {\rm tr}({\rm d}A)$. } 
\begin{equation}
    \delta {\cal O}_{\mathbb{P}^1}=\exp\left[-\int_{S^1\times M_4}\overline{\omega}\wedge {\rm tr}\left({\rm d}(h_k{\rm d}h_k^{-1}\wedge A)\right)\right].
\end{equation}
The integration over $M_4$ reduces to the pairing \eqref{pairing} and the remaining parts give a factor 
\begin{equation}
    \exp\left[q\int_{S^1}h_k{\rm d}h_k^{-1}\right]=e^{\frac{2\pi k}{N}q}.
\end{equation}
This gives a charge of ${\cal O}^d_{\mathbb{P}^1}$ that is consistent with that of the representation in \eqref{blowup-leading}.

\subsection{Universal expression for fundamental representations at one instanton}\label{sec:4.1}
In this section, we derive a universal expression for the one-instanton contribution to Wilson loop VEVs in fundamental representations of pure gauge theories. Here we call the representation with only the $i$-th Dynkin label 1 and all others 0 the $i$-th fundamental representation $\mathbf{r}_i$. 
Let $\omega^{\vee}=0,\mathbf{r}_{\rm{N}}=\mathbf{1}$ and $\mathbf{r}_{\rm{S}}=\mathbf{r}_i$, there exist blowup equations
\begin{equation}\label{eq:Zhatc1c2}
    \hat{Z}(b,\mathbf{1},\mathbf{r}_i)=c_1 Z_{{\bf r}_i}+c_2 Z_{\rm inst},\qquad \text{for } b=b_0-2,b_0,b_0+2,
\end{equation}
where $c_1=1$, $b_0=h_G^{\vee}\mod 2$ and $Z_{\rm inst}$ is the Nekrasov instanton partition function. For almost all cases, $c_2=0$. This is exactly the scheme proposed in \cite[(3.8)]{Kim:2021gyj}. 

In general, for rank-$r$ pure gauge theories, the K\"ahler parameters in the geometry can always be expressed as
\begin{align}
    t=({\alpha_1\cdot a,\cdots,\alpha_r\cdot a,\log{\frak{q}}+\sum_{i=1}^r d_i\,\alpha_i\cdot a}),
\end{align}
where $d_i$ are positive integers presented in the following table \footnote{The notation of simple roots is the same as the notation in the Mathematica package LieART \cite{Feger:2019tvk}.}:
\begin{align}
\begin{tabular}{|c|c|}
\hline
Gauge Algebra & $(d_i)$  \\
\hline
$A_1$ & $(1)$  \\
$A_{2r}$ & $(1,2,\cdots,r,r,\cdots,2,1)$  \\
$A_{2r-1}$ & $(1,2,\cdots,r,\cdots,2,1)$  \\
$B_2$ & $(1,1)$  \\
$B_{r}$ & $(2,4,6,\cdots,2r-4, 2r-3,2r-3)$  \\
$C_2$ & $(1,1)$  \\
$C_{r}$ & $(1,2,\cdots,r-1, \lfloor\frac{r+1}{2}\rfloor)$  \\
$D_{r}$ & $(2,4,6\cdots,2r-4,r-2,r-2)$  \\
$G_{2}$ & $(4,3)$  \\
$F_{4}$ & $(5,9,12,6)$  \\
$E_{6}$ & $(4, 8, 12, 8, 4, 6)$  \\
$E_{7}$ & $(8, 16, 24, 18, 12, 6, 12)$  \\
$E_{8}$ & $(20, 40, 60, 48, 36, 24, 12, 30)$  \\
\hline
\end{tabular}
\end{align}
The integers $d_i$ can be derived from the blowup equations, where the derivation will be presented in Appendix \ref{a:proof}. 

For any given representation ${\bf r}$, if its highest weight in the simple root basis is smaller than $(d_1,\cdots,d_r)$, then \eqref{eq:Zhatc1c2} holds and $c_2=0$; if it is equal to $(d_1,\cdots,d_r)$, $c_2\neq 0$ \footnote{The derivation for this fact will also be presented in Appendix \ref{a:proof}.}. For all pure gauge theories whose gauge algebras are simple Lie algebras, we have verified that all the fundamental representations ${\bf r}_i$ are lower representations of ${\bf r}_d$, except for $G=B_2$ and $\mathbf{r}=\mathbf{5}$. Then the VEV of Wilson loops can be solved from lower to higher instanton recursively from the strategies proposed in \cite{Kim:2019uqw}.

At one-instanton level, a universal expression can be derived. Utilizing the instanton expansion 
\begin{align}
    Z_{\rm inst}=1+Z^{\text{1-inst}}{\frak{q}}+\mathcal{O}(\frak{q}^2),\qquad Z_{{\bf r}}=\chi_{{\bf r}}+(\chi_{{\bf r}}Z^{\text{1-inst}}+\langle W_{{\bf r}}^{\text{1-inst}}\rangle)\mathfrak{q}+\mathcal{O}(\frak{q}^2),
\end{align}
and using the blowup equations \eqref{eq:Zhatc1c2}, we obtain 
\begin{align}\label{eq:Wr_one_inst}
    \langle W_{{\bf r}_i}^{\text{1-inst}}\rangle=-c^{\prime}+\sum_{\alpha_l\in \Delta_l}\frac{e^{(1-\lfloor{h_G^{\vee}}/{2} \rfloor)(\alpha_l\cdot a-\epsilon_1-\epsilon_2)}\sum_{\omega\in \mathbf{R}_i}e^{-\omega\cdot a}(q_1^{\alpha_l\cdot \omega}-1)(1-q_1)^{-1}(1-q_2)^{-1}}{(1-q_1q_2e^{-\alpha_l\cdot{a}})(1-e^{\alpha_l\cdot{a}})\prod_{\beta\cdot \alpha_l=-1}(1-e^{-\beta\cdot a})}.
\end{align}
Here $\mathbf{R}_i$ is the set of weights for the $i$-th fundamental representation $\mathbf{r}_i$. $c^{\prime}$ is related to $c_2^{(b)}$, which is non-zero only for $G=B_2$ and $r_i=\mathbf{5}$.  It can be determined by assuming 
\begin{align}
   \left. \langle W_{\mu_i}^{\text{1-inst}}\rangle \right|_{\alpha_i\cdot a\rightarrow \infty}=0.
\end{align}

Based on \eqref{eq:Wr_one_inst}, we bootstrap the VEVs $\langle W_{{\bf r}_i}^{\text{1-inst}}\rangle,\, i=1,\cdots,r,$ for all pure gauge theories with simple Lie gauge algebras \footnote{With the exception $\mu_3={\bf{6899079264}}$ and $\mu_4={\bf{146325270}}$ for $E_8$. We couldn't obtain the result in a reasonable amount of computation time.}.

For $A_r$ theories, we observed that
\begin{align}
    \langle W_{\mu_i}^{\text{1-inst}}\rangle= -v^{h_G^{\vee}}\sum_{l=0}^{\infty}\chi_{(\theta_{\mu_i}+l\,\theta_{G})} v^{2l},
\end{align}
where $\theta_{\mu_i}$ and $\theta_{G}$ stand for the highest weight for the $i$-th fundamental and adjoint representations of $G$. $\chi_{\theta}$ is the character of the representation with the highest weight $\theta$. This is the Hilbert-series-like expansion mentioned before. In other gauge groups, one can observe a similar expansion that only depends on $v$ for all the fundamental representations except for $G=E_8$. For a complete result of other representations by turning off all the gauge fugacities, see the supplementary {\it Mathematica} notebook uploaded along with the LaTeX files. 

\subsection{Other Examples}\label{s:rank-2}
In this subsection, we check the blowup equations \eqref{blowup-eq} explicitly for various 5d pure gauge theories with rank-two gauge groups $G$. In particular, we will focus on cases with non-trivial $\omega^{\vee}$. 
\subsubsection{SU(2)}

In the case of SU(2), the one-form symmetry is $\mathbb{Z}_2$ and the Wilson loop of the fundamental representation ${\bf 2}$ carries a non-trivial $\mathbb{Z}_2$ charge. Therefore, Wilson loops will have the same charge when they are in the tensor representation with the same number of fundamental representation mod 2. 

We focus on the case where $\omega^\vee$ is chosen as the fundamental coweight corresponding to the representation ${\bf 2}$, as the choice $\omega^\vee=0$ has been well studied in the literature (and is less interesting since there is no shift of the one-form symmetry charge between Wilson loops appearing in both sides of the blowup equation). The summation over ${n}$ appearing in the blowup equation \eqref{blowup-eq} is then shifted to a summation over a half-integer parameter $i\in\mathbb{Z}+\frac{1}{2}$. 

When $b=h^\vee=2$, the blowup equations are given by 
\begin{align}
    &\hat{Z}(2,{\bf 1},{\bf 1})=Z_{W_{\bf 1}}(a_1,\epsilon_1,\epsilon_2;\mathfrak{q})=Z^{\rm SU(2)}_{\rm inst}(a_1,\epsilon_1,\epsilon_2;\mathfrak{q})\label{eq:exam_su2_211},\\
    &q_1^{-\frac{1}{2}}\hat{Z}(2,{\bf 2},{\bf 1})=q_2^{-\frac{1}{2}}\hat{Z}(2,{\bf 1},{\bf 2})=Z_{W_{\bf 2}}(a_1,\epsilon_1,\epsilon_2;\mathfrak{q}).
\end{align}
Note that the representations $\mathbf{r}_{1,2}$ in the first equation \eqref{eq:exam_su2_211} are trivial, leading to the original blowup equation that appeared in \cite{Nakajima:2005fg,Nakajima:2003pg}.
When $b=4$, we have e.g. 
\begin{align}
    &\hat{Z}(4,{\bf 1},{\bf 1})=Z_{W_{\bf 2}}(a_1,\epsilon_1,\epsilon_2;\mathfrak{q}),\\
    &\hat{Z}(4,{\bf 2},{\bf 1})=q_1^{\frac{1}{2}}Z_{W_{{\bf 2}\otimes {\bf 2}}}(a_1,\epsilon_1,\epsilon_2;\mathfrak{q})-(q_1^{\frac{1}{2}}-q_1^{-\frac{1}{2}})(1-q_1\mathfrak{q})Z^{\rm SU(2)}_{\rm inst}(a_1,\epsilon_1,\epsilon_2;\mathfrak{q}),\\
    &\hat{Z}(4,{\bf 2}\otimes {\bf 2},{\bf 1})=q_1Z_{W_{{\bf 2}\otimes {\bf 2}\otimes {\bf 2}}}(a_1,\epsilon_1,\epsilon_2;\mathfrak{q})+2(1-q_1)(1-q_1\mathfrak{q})Z_{W_{{\bf 2}}}(a_1,\epsilon_1,\epsilon_2;\mathfrak{q}).
\end{align}
One can see that $\hat{Z}$ contains Wilson loops only with the representation of either an odd/even number of the fundamental representations ${\bf 2}$ tensored together, i.e. they have the same $\mathbb{Z}_2$ one-form symmetry charge. Similarly for $b=6$, 
\begin{align}
    \hat{Z}(6,{\bf 1},{\bf 1})=Z_{W_{{\bf 2}\otimes {\bf 2}}}(a_1,\epsilon_1,\epsilon_2;\mathfrak{q})-(1+(1-q_1-q_2-q_1q_2)\mathfrak{q}+q_1^2q_2^2\mathfrak{q}^2)Z^{\rm SU(2)}_{\rm inst}(a_1,\epsilon_1,\epsilon_2;\mathfrak{q}),
\end{align}
while for $b=0$, 
\begin{align}
    &\hat{Z}(0,{\bf 1},{\bf 1})=0,\\
    &\hat{Z}(0,{\bf 2},{\bf 1})=(q_1^{\frac{1}{2}}-q_1^{-\frac{1}{2}})Z^{\rm SU(2)}_{\rm inst}(a_1,\epsilon_1,\epsilon_2;\mathfrak{q}).
\end{align}
We remark that all the blowup equations presented above are found through explicit computations at low orders and then checked at least up to the 6-instanton level. 

\subsubsection{SU(3)}

SU(3) gauge theory has a $\mathbb{Z}_3$ one-form symmetry, and it gives the most interesting examples among rank-2 gauge theories. The Wilson loop of the fundamental representation ${\bf 3}$ carries one-form symmetry charge $1$, and that of the anti-fundamental representation $\bar{\bf 3}$ bears a charge of $2$. This brings us a more abundant structure of linear combinations of Wilson loops with the same charge appearing in the blowup equation. 

Let $\omega_1$ and $\omega_2$ be the fundamental weights corresponding to the representations ${\bf 3}$ and $\bar{\bf 3}$. Then for the choice of $\omega^\vee=\omega^\vee_1$, we have e.g. 
\begin{align}
    &\hat{Z}(b=3,{\bf 1},{\bf 1})=Z^{\rm SU(3)}_{\rm inst},\quad
    \hat{Z}(b=5,{\bf 1},{\bf 1})=Z_{W_{{\bf 3}}},\quad 
    \hat{Z}(b=-5,{\bf 1},{\bf 1})=Z_{W_{\bar{\bf 3}}},
\end{align}
and 
\begin{align}
    &\hat{Z}(b=5,{\bf 1},{\bf 3})=q_2^{\frac{2}{3}}\lt(Z_{W_{{\bf 3}\otimes {\bf 3}}}-(1-q_2^{-1})Z_{W_{\bar{\bf 3}}}\rt),\\
    &\hat{Z}(b=-5,{\bf 1},{\bf 3})=q_2^{-\frac{1}{3}}\lt(Z_{W_{{\bf 3}\otimes \bar{\bf 3}}}-(1-q_2-q_1^{-\frac{1}{2}}q_2^{-\frac{1}{2}}(1-q_2^{-1})\mathfrak{q})Z^{\rm SU(3)}_{\rm inst}\rt).
\end{align}
It is clear from the above that $\hat{Z}$ is a linear combination of Wilson loops with the same charge. When $|b|<h^\vee$, we have 
\begin{align}
    &\hat{Z}(b=1,{\bf 1},{\bf 3})=(1-q_2^{-1})Z^{\rm SU(3)}_{\rm inst},\quad \hat{Z}(b=-1,{\bf 1},{\bf 3})=0,
\end{align}
which indicates that $\hat{Z}$ only contains Wilson loops of representation with lower weights, and in particular for $b=-1$, there is no lower-weight representation than ${\bf 3}$ with charge $2$, so $\hat{Z}$ becomes zero. 

Similarly for $\omega^\vee=\omega_2^\vee$, 
\begin{align}
    &\hat{Z}(b=3,{\bf 1},{\bf 1})=Z^{\rm SU(3)}_{\rm inst},\quad
    \hat{Z}(b=5,{\bf 1},{\bf 1})=Z_{W_{\bar{\bf 3}}},
\end{align}
and 
\begin{align}
    &\hat{Z}(b=5,{\bf 1},{\bf 3})=q_2^{\frac{1}{3}}(Z_{{\bf 3}\otimes \bar{\bf 3}}+(1-q_2)(q_2^{-1}+q_1^{\frac{1}{2}}q_2^{\frac{1}{2}}\mathfrak{q})Z^{\rm SU(3)}_{\rm inst}).
\end{align}
All the blowup equations presented here are checked up to the 3-instanton level.

\subsubsection{SO(5)}\label{s:blowup-so5}

The one-form symmetry in $\mathfrak{so}(5)$ gauge theory\footnote{As stated before, the correct description for the theory is an Spin(5) gauge theory.} is $\mathbb{Z}_2$, and it matches with the quotient of the coweight lattice by the coroot lattice. This means that we only have two choices for $\omega^\vee$, either $\omega^\vee=0$ or $\omega^\vee=\omega_1^\vee$, where $\omega_1$ stands for the fundamental weight corresponding to the vector representation ${\bf 5}$. In this section, we present the blowup equations associated with $\omega^\vee=\omega_1^\vee$, but we note that the one-form symmetry charge is always equal between the Wilson loops appearing in the l.h.s. of the blowup equation \eqref{blowup-eq}, $\hat{Z}$, and those in the r.h.s., as the non-trivial $\mathbb{Z}_2$ charge is carried by the spinor representation ${\bf 4}$. 

We have, for example, 
\begin{align}
    \hat{Z}(3,{\bf 1},{\bf 1})&=Z^{\rm SO(5)}_{\rm inst},\quad \hat{Z}(5,{\bf 1},{\bf 1})=Z_{W_{\bf 5}}-q_1^{\frac{1}{2}}q_2^{\frac{1}{2}}\mathfrak{q}Z^{\rm SO(5)}_{\rm inst},\\
    \hat{Z}(\pm 1,{\bf 1},{\bf 1})&=0,\quad \hat{Z}(-5,{\bf 1},{\bf 1})=-Z_{W_{\bf 5}}+q_1^{-\frac{1}{2}}q_2^{-\frac{1}{2}}\mathfrak{q}Z^{\rm SO(5)}_{\rm inst},\\
     \hat{Z}(\pm3,{\bf 1},{\bf 4})&=q_2^{\pm\frac{1}{2}}Z_{W_{{\bf 4}}},\quad \hat{Z}(\pm1,{\bf 1},{\bf 4})=0,\\
    \hat{Z}(7,{\bf 1},{\bf 1})&=Z_{W_{{\bf 5}\otimes {\bf 5}}}-Z_{W_{{\bf 4}\otimes {\bf 4}}}+(1+q_1^{\frac{1}{2}}q_2^{\frac{1}{2}}(1+q_1+q_2)\mathfrak{q})Z_{W_{{\bf 5}}}\cr
    &+(q_1^{\frac{1}{2}}q_2^{\frac{1}{2}}\mathfrak{q}-q_1q_2(q_1+q_2)\mathfrak{q}^2)Z^{\rm SO(5)}_{\rm inst},\\
    \hat{Z}(5,{\bf 1},{\bf 4})&=q_2^{\frac{1}{2}}Z_{W_{{\bf 5}\otimes {\bf 4}}}+\left[(q_2^{-\frac{1}{2}}-q_2^{\frac{1}{2}})-q_1^{\frac{1}{2}}q_2^{\frac{3}{2}}\mathfrak{q}\right]Z_{W_{{\bf 4}}},\\
    \hat{Z}(-5,{\bf 1},{\bf 4})&=-q_2^{-\frac{1}{2}}Z_{W_{{\bf 5}\otimes {\bf 4}}}+\left[(q_2^{\frac{1}{2}}-q_2^{-\frac{1}{2}})-q_1^{-\frac{1}{2}}q_2^{-\frac{3}{2}}\mathfrak{q}\right]Z_{W_{{\bf 4}}}.
\end{align}
We checked them up to three instantons. Note that $h^\vee=3$ in this example. 

\subsubsection{$G_2$}\label{s:blowup-G2}

There is no non-trivial one-form symmetry in $G_2$ gauge theory, so all Wilson loops with representations of lower weight than \eqref{blowup-leading} appear in $\hat{Z}$. Note that $h^\vee=4$ for $G_2$, and we have for example, 
\begin{align}
    &\hat{Z}(\pm 4,{\bf 1},{\bf 1})=Z^{{\rm G}_2}_{\rm inst},\quad \hat{Z}(\pm 4,{\bf 7},{\bf 1})=Z_{W_{\bf 7}},\quad \hat{Z}(\pm 2,{\bf 14},{\bf 1})=Z_{W_{\bf 14}},\\
    &\hat{Z}(\pm 4,{\bf 14},{\bf 1})=Z_{W_{\bf 14}}+(q_1^{\pm 2}-q_1^{\pm 3})q_2^{\pm1}\mathfrak{q}Z^{{\rm G}_2}_{\rm inst},\\
    &\hat{Z}(\pm 6,{\bf 1},{\bf 1})=(1-q_1^{\pm2}q_2^{\pm2}\mathfrak{q})Z^{{\rm G}_2}_{\rm inst},\quad \hat{Z}(\pm 6,{\bf 7},{\bf 1})=(1-q_1^{\pm3}q_2^{\pm2}\mathfrak{q})Z_{W_{\bf 7}},\\
    &\hat{Z}(\pm8,{\bf 1},{\bf 1})=\left(1-(q_1^{\pm2}q_2^{\pm2}+q_1^{\pm3}q_2^{\pm2}+q_1^{\pm2}q_2^{\pm3})\mathfrak{q}\right)Z^{{\rm G}_2}_{\rm inst}-q_1^{\pm3}q_2^{\pm3}\mathfrak{q}Z_{W_{\bf 14}}.
\end{align}
We checked these blowup equations up to three instantons.

\section{Conclusion and Discussion}\label{s:sum}

In this article, we presented a comprehensive investigation into Wilson loops in 5d ${\cal N}=1$ pure gauge theories, encompassing both classical and exceptional gauge groups. In particular, a survey is conducted on higher-rank theories with Wilson loops in higher representations. The main results include the systematic formulation of general blowup equations for Wilson loops, leveraging constraints derived from one-form symmetry and instanton partition functions at low instanton levels. The consistency of these equations was rigorously verified through explicit computations using both Chern-character insertions and qq-characters, applied to a wide range of representations and gauge groups. 

Furthermore, we observed a universal Hilbert series-like structure in a family of one-instanton Wilson loop free energies, reminiscent of instanton partitions in pure gauge theories. This universality across different gauge groups and representations warrants further investigation.

There are several future directions to be explored in relation to this work. A central insight from this work is the pivotal role played by one-form symmetry in the construction/restriction of the codimension-4 defect partition functions, which combine into the qq-characters. The qq-characters, on the other hand, serve as generators of the dual quiver ${\cal W}$-algebra \cite{Nekrasov:2015wsu,Kimura:2015rgi,Bourgine:2015szm,Bourgine:2016vsq}, and it is curious whether the one-form symmetry or its generalization can be related to some symmetry acting on the expansion modes in 2d CFT. The same question might also be asked in the context of Alday-Gaiotto-Tachikawa (AGT) relation \cite{Alday:2009aq,Wyllard:2009hg}, which describes the dual 2d CFT (${\cal W}$-algebra) in an ${\cal S}$-dual picture.


One more interesting direction to explore is the so-called Urod algebra \cite{Bershtein:2013oka} hidden behind the blowup equations. It combines the AGT relation and the blowup equation to establish an isomorphism between the tensor product of conformal vertex algebras (the ${\cal W}$-algebra in AGT sense) ${\cal M}_{b_1}\otimes {\cal M}_{b_2}$ (with $b_1=b/\sqrt{b^2-1}$, $b_2=\sqrt{1-b^2}$ and $b=\sqrt{\epsilon_1/\epsilon_2}$), which comes from the r.h.s. of \eqref{blowup-eq}, and an algebra ${\cal A}={\cal M}_b\otimes {\cal U}$ associated to the l.h.s. ${\cal U}$ is a vertex algebra called the Urod algebra mentioned before. In the case of SU(2) gauge group, it is identified as a deformation of $\hat{\mathfrak{sl}}(2)_1$ containing two commuting Virasoro subalgebras with central charges $c=-\frac{22}{5}$ and $c=-\frac{3}{5}$, and its generalization to higher-rank cases is discussed in \cite{Arakawa:2020oqo}. It remains unclear how to explicitly utilize the Urod algebra to derive/prove the blowup equation, and it will be more exciting if one can use the Urod algebra to explain the coefficients appearing in the blowup equations for Wilson loops, which cannot be determined purely from the one-form symmetry. 

Finally, the robust tools established in this work for computing Wilson loops are expected to significantly contribute to the understanding of five-dimensional supergravity in future works.

\acknowledgments

We would like to thank Jie Gu and Jiahua Tian for their helpful discussion. XW is supported by the Fundamental Research Funds for the Central Universities (Grants No. WK2030250140) and the National Natural Science Foundation of China (Grant No.12247103). R.Z. was partially supported by the National Natural Science Foundation of China (Grant No.~12105198). 

\appendix

\section{1-loop determinants for Sp($N$)}\label{a:Sp}

This appendix compiles the 1-loop determinants for Sp($N$) theories, which serve as integrands for computing instanton partition functions. 

We adopt the following notations: $\sh (x):= 2\sinh\left(x/2\right)$, $\ch (x):= 2\cosh\left(x/2\right)$. A bracket with a $\pm$ sign is a short-hand for a product: $[\pm x+y]\equiv[x+y][-x+y]$. The instanton number $k$ is parametrized as $k=2n+\chi$ with $\chi=0,1$. 

The instanton partition function \eqref{Sp-integral} receives contributions from vector multiplets and defect insertions. The 1-loop determinants for the vector multiplets are given by: 
\begin{align}
    Z^{\left(k=2n+\chi,+\right)}_{\text{vec}}&=\prod_{I=1}^n \left(\frac{\sh\left(\pm\phi_I\right) \sh (\pm\phi_I+2\epsilon_+)}{\sh\left(\pm\phi_I \pm \epsilon_{-}+\epsilon_{+}\right)}\right)^{\chi} \prod_{I<J}^{n}\frac{\sh\left(\pm\phi_I\pm\phi_J\right) \sh(\pm\phi_I\pm\phi_J+2\epsilon_+)}{\sh\left(\pm\phi_I\pm\phi_J\pm \epsilon_{-}+\epsilon_{+}\right)}\notag
    \\
    &\times\left(\prod_{I=1}^n \frac{1}{\sh\left(\pm2\phi_I \pm \epsilon_{-}+\epsilon_{+}\right)}\right)\frac{\sh\left(2\epsilon_{+}\right)^n}{\sh\left(\pm \epsilon_{-}+\epsilon_{+}\right)^{n+\chi}}\notag
    \\
    &\times\prod_{i=1}^N \frac{1}{\sh\left(\pm a_i+\epsilon_{+}\right)^{\chi}}\prod_{I=1}^n \frac{1}{\sh\left(\pm \phi_I \pm a_i + \epsilon_{+}\right)},
\end{align}
\begin{align}
    Z^{\left(k=2n+1,-\right)}_{\text{vec}}&=\prod_{I=1}^n \frac{\text{ch}(\pm\phi_I)\text{ch}(\pm\phi_{I}+2\epsilon_{+})}{\text{ch}(\pm\phi_I\pm\epsilon_-+\epsilon_+)\sh(\pm2\phi_I\pm\epsilon_-+\epsilon_+)}\notag
    \\
    &\times\left(\prod_{I<J}^n\frac{\sh(\pm\phi_I\pm\phi_J)\sh(\pm\phi_{I}\pm\phi_{J}+2\epsilon_{+})}{\sh(\pm\phi_I\pm\phi_J\pm\epsilon_-+\epsilon_+)}\right)\frac{\sh(2\epsilon_+)^n}{\sh(\pm\epsilon_-+\epsilon_+)^{n+1}}\notag
    \\
    &\times\prod_{i=1}^N \frac{1}{\text{ch}\left(\pm a_i+\epsilon_{+}\right)}\prod_{I=1}^n \frac{1}{\sh\left(\pm \phi_I \pm a_i + \epsilon_{+}\right)},
\end{align}
and
\begin{align}
    Z^{\left(k=2n,-\right)}_{\text{vec}}&=\prod_{I=1}^{n-1}\frac{\sh(\pm2\phi_I)\sh(\pm2\phi_I+4\epsilon_+)}{\sh(\pm2\phi_I\pm2\epsilon_-+2\epsilon_+)\sh(\pm2\phi_I\pm\epsilon_-+\epsilon_+)}\notag
    \\
    &\times\left(\prod_{I<J}^{n-1}\frac{\sh(\pm\phi_I\pm\phi_J)\sh(\pm\phi_I\pm\phi_J+2\epsilon_+)}{\sh(\pm\phi_I\pm\phi_J\pm\epsilon_-+\epsilon_+)}\right)\frac{\text{ch}(2\epsilon_+)\sh(\epsilon_+)^{n-1}}{\sh(\pm2\epsilon_-+2\epsilon_+)\sh(\pm\epsilon_-+\epsilon_+)^n}\notag
    \\
    &\times\prod_{i=1}^N \frac{1}{\sh\left(\pm 2 a_i+2 \epsilon_{+}\right)}\prod_{I=1}^{n-1} \frac{1}{\sh\left(\pm \phi_I \pm a_i + \epsilon_{+}\right)}.
\end{align}
The defect contributions are given by
\begin{align}
    Z^{\left(k=2n+\chi,+\right)}_{\text{defect}}&=\left(\frac{\sh(\pm X-\epsilon_-)}{\sh(\pm X-\epsilon_+)}\right)^\chi\prod_{I=1}^n\frac{\sh(\pm\phi_I\pm X-\epsilon_-)}{\sh(\pm\phi_I\pm X-\epsilon_+)},
    \\
    Z^{\left(k=2n+1,-\right)}_{\text{defect}}&=\frac{\text{ch}(\pm X-\epsilon_-)}{\text{ch}(\pm X-\epsilon_+)}\prod_{I=1}^n\frac{\sh(\pm\phi_I\pm X-\epsilon_-)}{\sh(\pm\phi_I\pm X-\epsilon_+)},
    \\
    Z^{\left(k=2n,-\right)}_{\text{defect}}&=\frac{\sh(\pm2X-2\epsilon_-)}{\sh(\pm2X-2\epsilon_+)}\prod_{I=1}^{n-1}\frac{\sh(\pm\phi_I\pm X-\epsilon_-)}{\sh(\pm\phi_I\pm X-\epsilon_+)}.
\end{align}

\section{Explicit Expressions}

In this appendix, we further give explicit expressions of Wilson loops to check the isomorphism $\mathfrak{su}(4)\cong\mathfrak{so}(6)$ and $\mathfrak{su}(2)\cong\mathfrak{sp}(1)$ for higher-representations beyond the fundamental representation. Although the equalities are expected to hold trivially, they are not that obvious at the level of concrete computation. The consistency check also guarantees the validness of the Chern-character insertion method described in section \ref{s:Chern}.

\subsection{SU(4) vs SO(6)}\label{a:su4}

In addition to the rank-2 examples shown in section \ref{s:rank-2}, we also checked some equivalence relations induced from the Lie-algebra isomorphism to further support the prescriptions to compute the Wilson loops described in Section \ref{s:Chern} and Section \ref{s:qq}. In this section, we examine the equivalence between the Wilson loops in isomorphic representations in SU(4) and SO(6) theories.

To begin, we utilize the qq-character approach. Considering the instanton partition function with one defect, 
\begin{align}
    &\frac{Z^{\rm SU(4)}_{\text{1 defect}}(x)}{Z^{\rm SU(4)}_{\rm inst}}=1\cdot x^{-2}-W_{\mathbf{4}}\cdot x^{-1}+W_{\mathbf{6}}\cdot x^{0}-W_{\mathbf{\Bar{4}}}\cdot x+1\cdot x^2,\\
    &\frac{Z^{\rm SO(6)}_{\text{1 defect}}(x)}{Z^{\rm SO(6)}_{\rm inst}}=1\cdot x^{-3}-W_{\mathbf{6}}\cdot x^{-2}+W_{\mathbf{15}}\cdot x^{-1}-W_{\mathbf{20}}+W_{\mathbf{15}}\cdot x-W_{\mathbf{6}}\cdot x^{2}+1\cdot x^{3}.\label{qq-so6-1}
\end{align}
Representation $\mathbf{6}$ denotes the second antisymmetric tensor product of the fundamental representation of $\mathfrak{su}(4)$, which can be compared with the vector representation of $\mathfrak{so}(6)$. While reference \cite{Nawata:2023wnk} examined this equivalence in the unrefined limit up to the 4-instanton level, we have verified it up to 3 instantons on the generic $\Omega$-background, i.e. 
\begin{align}
    \left.\frac{Z_{\rm 1 defect}^{\rm SU(4)}}{Z^{\rm SU(4)}_{\rm inst}}\right|_{x^0}=W_{\mathbf{6}}=-\left.\frac{Z_{\rm 1 defect}^{\rm SO(6)}}{Z^{\rm SO(6)}}\right|_{x^2}.
\end{align}

A similar comparison applies to two-defect partition functions containing Wilson loops of higher-weight representations. For instance, the $\mathbf{4}\otimes\mathbf{\Bar{4}}$ Wilson loop can be obtained from the two-defect partition function in SU(4) theory, and it is decomposed into irreducible representations as $\mathbf{4}\otimes\mathbf{\Bar{4}}={\bf 15}\oplus {\bf 1}$. It can then be compared with $W_{\mathbf{15}}$ appearing in the one-defect partition function of SO(6), \eqref{qq-so6-1}. The expansion of the two-defect partition function in SU(4) is given in terms of the Wilson loops as 
\begin{align}
    \frac{Z^{\rm SU(4)}_{\text{2 defect}}(x_1,x_2)}{Z^{\rm SU(4)}_{\rm inst}}&=1\cdot\left( x_1^{-2}x_2^{-2}+x_1^{2}x_2^{-2} +x_1^{-2}x_2^{2}+ x_1^{2}x_2^{2}\right)\notag
    \\
    &-W_{\mathbf{4}}\cdot\left( x_1^{-1}x_2^{-2} +x_1^{-2}x_2^{-1}+ x_1^{2}x_2^{-1}+ x_1^{-1}x_2^{2}\right)\notag
    \\
    &+W_{\mathbf{6}}\cdot\left(x_1^{0}x_2^{-2}+x_1^{-2}x_2^{0}+ x_1^{2}x_2^{0}+ x_1^{0}x_2^{2}\right)\notag
    \\
    &-W_{\mathbf{\Bar{4}}}\cdot\left( x_1^{1}x_2^{-2}+x_1^{-2}x_2^{1}+x_1^{2}x_2^{1}+x_1^{1}x_2^{2}\right)\notag
    \\
    &+W_{\mathbf{4}\otimes\mathbf{4}}\cdot x_1^{-1}x_2^{-1}-W_{\mathbf{4}\otimes\mathbf{6}}\cdot\left(x_1^{-1}x_2^{0}+x_1^{0}x_2^{-1}\right)+W_{\mathbf{4}\otimes\mathbf{\Bar{4}}}\cdot\left( x_1^{-1}x_2^{1}+x_1^{1}x_2^{-1}\right)\notag
    \\
    &+W_{\mathbf{6}\otimes\mathbf{6}}\cdot x_1^{0}x_2^{0}-W_{\mathbf{6}\otimes\mathbf{\Bar{4}}}\cdot\left( x_1^{1}x_2^{0}+ x_1^{0}x_2^{1}\right)+W_{\mathbf{\Bar{4}}\otimes\mathbf{\Bar{4}}}\cdot x_1^{1}x_2^{1}\notag
    \\
    &+\mathfrak{q}\frac{(1-q_1) (1-q_2)(1+q_1q_2)x_1x_2}{(x_1-x_2q_1q_2)(x_2-x_1q_1q_2)}.
\end{align}
Among the expansion coefficients, $W_{\mathbf{6}\otimes\mathbf{6}}$ also appears, which can be compared to the Wilson loop of the same representation in the two-defect partition function of SO(6), thereby providing another non-trivial check of the isomorphism.
\begin{align}
    \frac{Z^{\rm SO(6)}_{\text{2 defect}}(x_1,x_2)}{Z^{\rm SO(6)}_{\rm inst}}&=1\cdot\left(x_1^{-3}x_2^{-3}+x_1^{-3}x_2^{3}+x_1^{3}x_2^{-3}+x_1^{3}x_2^{3}\right)\notag
    \\
    &-W_{\mathbf{6}}\cdot\left(x_1^{-3}x_2^{-2}+x_1^{-3}x_2^{2}+x_1^{-2}x_2^{-3}+x_1^{-2}x_2^{3}+x_1^{2}x_2^{-3}+x_1^{2}x_2^{3}+x_1^{3}x_2^{-2}+x_1^{3}x_2^{2}\right)\notag
    \\
    &+W_{\mathbf{15}}\cdot\left(x_1^{-3}x_2^{-1}+x_1^{-3}x_2^{1}+x_1^{-1}x_2^{-3}+x_1^{-1}x_2^{3}+x_1^{1}x_2^{-3}+x_1^{1}x_2^{3}+x_1^{3}x_2^{-1}+x_1^{3}x_2^{1}\right)\notag
    \\
    &-W_{\mathbf{20}}\cdot\left(x_1^{-3}x_2^{0}+x_1^{0}x_2^{-3}+x_1^{0}x_2^{3}+x_1^{3}x_2^{0}\right)\notag
    \\
    &+W_{\mathbf{6}\otimes\mathbf{6}}\cdot\left(x_1^{-2}x_2^{-2}+x_1^{-2}x_2^{2}+x_1^{2}x_2^{-2}+x_1^{2}x_2^{2}\right)\notag
    \\
    &-W_{\mathbf{6}\otimes\mathbf{15}}\cdot\left(x_1^{-2}x_2^{-1}+x_1^{-2}x_2^{1}+x_1^{-1}x_2^{-2}+x_1^{-1}x_2^{2}+x_1^{1}x_2^{-2}+x_1^{1}x_2^{2}+x_1^{2}x_2^{-1}+x_1^{2}x_2^{1}\right)\notag
    \\
    &+W_{\mathbf{6}\otimes\mathbf{20}}\cdot\left(x_1^{-2}x_2^{0}+x_1^{0}x_2^{-2}+x_1^{0}x_2^{2}+x_1^{2}x_2^{0}\right)\notag
    \\
    &+W_{\mathbf{15}\otimes\mathbf{15}}\cdot\left(x_1^{-1}x_2^{-1}+x_1^{-1}x_2^{1}+x_1^{1}x_2^{-1}+x_1^{1}x_2^{1}\right)\notag
    \\
    &-W_{\mathbf{15}\otimes\mathbf{20}}\cdot\left(x_1^{-1}x_2^{0}+x_1^{0}x_2^{-1}+x_1^{0}x_2^{1}+x_1^{1}x_2^{0}\right)\notag
    \\
    &+W_{\mathbf{20}\otimes\mathbf{20}}\cdot x_1^{0}x_2^{0}+\mathfrak{q}\cdot   Z_{\text{Extra}},
\end{align}
where 
\begin{align}
    Z_{\text{Extra}}&= \frac{1}{ x_1^2 x_2^2 }\Bigg( \frac{1}{q_1 q_2\left(x_1x_2-q_1q_2\right) \left(x_1-x_2q_1q_2\right) \left(x_2-x_1q_1q_2\right) \left(1-x_1x_2q_1q_2\right)}\notag
    \\
    &\times \Big(4\left(1+x_1^4x_2^4\right)\left(x_1^4+x_2^4\right)q_1^3q_2^3 +2x_1^4x_2^4\left(1+q_2+q_2^2+q_2^3+q_1\left(1+8q_2+4q_2^2-2q_2^3\right.\right. \notag
    \\
    &\left.+q_2^4\right) +q_1^2\left(1+4q_2-17q_2^2+q_2^3+6q_2^4+q_2^5\right)+q_1^3\left(1-2q_2+q_2^2+32q_2^3+q_2^4-2q_2^5\right.\notag
    \\
    &\left.+q_2^6\right)+q_1^4q_2\left(1+6q_2+q_2^2-17q_2^3+4q_2^4+q_2^5\right)+q_1^5q_2^2\left(1-2q_2+4q_2^2+8q_2^3+q_2^4\right)\notag
    \\
    &\left. +q_1^6q_2^3\left(1+q_2+q_2^2+q_2^3\right)\right) +2x_1^2x_2^2\left(1+x_1^2x_2^2\right)\left(x_1^2+x_2^2\right)q_1q_2\left(2+q_2-q_2^2 \right.\notag
    \\
    &+q_1\left(1-8q_2+q_2^2+2q_2^3\right)+q_1^2\left(-1+q_2+16q_2^2+q_2^3-q_2^4\right)+q_1^3q_2\left(2+q_2-8q_2^2\right.\notag
    \\
    &\left.\left.+q_2^3\right)+q_1^4q_2\left(-1+q_2+2q_2^2\right)\right)+x_1^2x_2^2\left(1+x_1^4\right)\left(1+x_2^4\right)q_1q_2\left( -1+q_2+q_1\left(1-10q_2\right.\right. \notag
    \\
    &\left.\left.-q_2^2+2q_2^3\right)-q_1^2q_2\left(1-10q_2+q_2^2\right)+q_1^3q_2\left(2-q_2-10q_2^2+q_2^3\right)-q_1^4q_2^3\left(-1+q_2\right)\right)\notag
    \\
    &-x_1^3x_2^3\left(1+x_1^2\right)\left(1+x_2^2\right)q_1q_2\left(-5-q_2+5q_2^2+q_2^3+q_1\left(-1+27q_2+4q_2^2-3q_2^3\right.\right.\notag
    \\
    &\left.+q_2^4\right)+q_1^2\left(5+4q_2-26q_2^2+4q_2^3+5q_2^4\right)+q_1^3\left(1-3q-2+4q_2^2+27q_2^3-q_2^4\right)\notag
    \\
    &\left.-q_1^4q_2\left(-1-5q_2+q_2^2+5q_2^3\right)\right)+x_1x_2\left(1+x_1^2\right)\left(1+x_2^2\right)q_1^2q_2^2\left(\left(1+x_1^4\right)\left(1+x_2^4\right)\right.\notag
    \\
    &\left(1-q_1\right)\left(1-q_2\right)\left(1+q_1q_2\right)+\left(x_1^2+x_2^2-x_1^2x_2^2+x_1^4x_2^2+x_1^2x_2^4\right)\left(-5+8q_1q_2+q_2^2\right.\notag
    \\
    &\left.\left.+q_1^2\left(1-5q_2^2\right)\right)\right)\Big)-8x_1x_2\left(1+x_1^2\right)\left(1+x_2^2\right)\Bigg).
\end{align}

To further validate the isomorphism, we have examined the equivalence of Wilson loops in the ${\bf 15}$ and ${\bf 6}\otimes {\bf 6}$ representations between SU(4) and SO(6) theories up to three-instanton level. The following relations hold: 
\begin{align}
    W_{\mathbf{6}\otimes\mathbf{6}}&=\left.\left(\frac{Z_{\text{2\,defect}}^{\rm SU(4)}}{Z^{\rm SU(4)}}-\mathfrak{q}\frac{(1-q_1) (1-q_2)(1+q_1q_2)x_1x_2}{(x_1-x_2q_1q_2)(x_2-x_1q_1q_2)}\right)\right|_{x_1^0,x_2^0}\notag
    \\
    &=\left.\left(\frac{Z_{\text{2\,defect}}^{\rm SO(6)}}{Z^{\rm SO(6)}}-\mathfrak{q}\cdot Z_{\text{Extra}}\right)\right|_{x_1^{-2}x_2^{-2}},
    \\
    W_{\mathbf{4}\otimes\mathbf{\Bar{4}}}&=W_{\mathbf{15}}+1=\left.\left(\frac{Z_{\text{2\,defect}}^{\rm SU(4)}}{Z^{\rm SU(4)}}-\mathfrak{q}\frac{(1-q_1) (1-q_2)(1+q_1q_2)x_1x_2}{(x_1-x_2q_1q_2)(x_2-x_1q_1q_2)}\right)\right|_{x_1^{1}x_2^{-1}}\notag
    \\
    &=\left.\frac{Z_{\text{1\,defect}}^{\rm SO(6)}}{Z^{\rm SO(6)}}\right|_{x^{1}}+1 .
\end{align}
We now present the explicit one-instanton corrections for Wilson loops in representations ${\bf 6}$, ${\bf 4}\otimes \bar{\bf 4}$, and ${\bf 6}\otimes {\bf 6}$. These results are expressed in terms of characters of $\mathfrak{su}(4)$, which is isomorphic to $\mathfrak{so}(6)$.

With the notation $v:=\sqrt{q_1q_2}$, we have 
\begin{align}
    F_{\mathbf{6}}^{(1)}&=\frac{v^4}{\prod_{\alpha\in \Delta}\left(1-e^{-\vec{\alpha}\cdot \vec{a}}v^2\right)}\cdot\Big(v^8\chi_{\mathbf{300}}-\left(v^6+v^8+v^{10}\right)\chi_{\mathbf{126}}-\left(v^6+v^8+v^{10}\right)\chi_{\mathbf{\overline{126}}}\notag
   \\
   &+v^8\chi_{\mathbf{70}}+v^8\chi_{\mathbf{\overline{70}}}-2\left(v^6+v^{10}\right)\chi_{\mathbf{64}}+\left(v^4+v^6+v^8\right)^2\chi_{\mathbf{\overline{50}}}\notag
   \\  
   &+\left(v^2+2v^4+2v^6+3v^8+2v^{10}+2v^{12}+v^{14}\right)\chi_{\mathbf{10}}\notag
   \\
   &+\left(v^2+2v^4+2v^6+3v^8+2v^{10}+2v^{12}+v^{14}\right)\chi_{\mathbf{\overline{10}}}\notag
   \\
   &-\left(1+2v^2+3v^4+4v^6+3v^8+4v^{10}+3v^{12}+2v^{14}+v^{16}\right)\chi_{\mathbf{6}}\Big),
 \end{align} 
 \begin{align}
    F_{\mathbf{4}\otimes\mathbf{\Bar{4}}}^{(1)}&=\frac{\left(1-q_1\right)\left(1-q_2\right)v^2}{\prod_{\alpha\in \Delta}\left(1-e^{-\vec{\alpha}\cdot \vec{a}}v^2\right)}\cdot\Big(-\left(v^8+v^{10}\right)\chi_{\mathbf{84}}+\left(v^6+v^8+v^{10}+v^{12}\right)\chi_{\mathbf{45}}\notag
    \\
    &+\left(v^6+v^8+v^{10}+v^{12}\right)\chi_{\overline{\mathbf{45}}}-\left(v^4+v^6+v^8+v^{10}+v^{12}+v^{14}\right)\chi_{\mathbf{20'}}\notag
    \\
    &-\left(v^4+v^6+2v^8+2v^{10}+v^{12}+v^{14}\right)\chi_{\mathbf{15}}\notag
    \\
    &+\left(1+3v^2+5v^4+7v^6+8v^8+8v^{10}+7v^{12}+5v^{14}+3v^{16}+v^{18}\right)\chi_{\mathbf{1}}\Big),
\end{align}
\begin{align}
    F_{\mathbf{6}\otimes\mathbf{6}}^{(1)}&=\frac{\left(1-q_1\right)\left(1-q_2\right)\left(1-v^2\right)}{\prod_{\alpha\in \Delta}\left(1-e^{-\vec{\alpha}\cdot \vec{a}}v^2\right)}\cdot\Big(-v^{10}\chi_{\mathbf{300'}}+\left(v^8+v^{12}\right)\chi_{\mathbf{256}}+\left(v^8+v^{12}\right)\chi_{\overline{\mathbf{256}}}\notag
    \\
    &-\left(v^6+2v^{10}+v^{14}\right)\chi_{\mathbf{175}}-\left(v^6-v^8+3v^{10}-v^{12}+v^{14}\right)\chi_{\mathbf{84}}\notag
    \\
    &+\left(v^4+4v^8+4v^{12}+v^{16}\right)\chi_{\mathbf{45}}+\left(v^4+4v^8+4v^{12}+v^{16}\right)\chi_{\overline{\mathbf{45}}}\notag
    \\
    &-\left(v^6+v^8+2v^{10}+v^{12}+v^{14}\right)\chi_{\mathbf{35}}-\left(v^6+v^8+2v^{10}+v^{12}+v^{14}\right)\chi_{\overline{\mathbf{35}}}\notag
    \\
    &+\left(v^4-v^6+2v^8-2v^{10}+2v^{12}-v^{14}+v^{16}\right)\chi_{\mathbf{20'}}\notag
    \\
    &-\left(v^2+v^4+5v^6+v^8+8v^{10}+v^{12}+5v^{14}+v^{16}+v^{18}\right)\chi_{\mathbf{15}}\notag
    \\
    &+\left(1+3v^2+6v^4+8v^6+11v^8+10v^{10}+11v^{12}+8v^{14}+6v^{16}+3v^{18}+v^{20}\right)\chi_{\mathbf{1}}\Big).
\end{align}

Our calculations reveal a key structural feature, analogous to that noted in \eqref{1-ins-Wilson}: up to an overall factor of ${\cal I}^{|{\rm q}|-1}$ defined in \eqref{def-I}, the (normalized) one-instanton Wilson loop depend solely on $v$ (i.e. $q_1q_2=:v^2$) and not on the ratio $q_1/q_2$. In the absence of Wilson loops, this property finds a natural explanation within the Hilbert series framework, as the instanton partition function itself has been identified with a Hilbert series under a suitable map \cite{Rodriguez-Gomez:2013dpa}. Our findings thus suggest a potential extension of this connection, linking the Hilbert series to Wilson loops observables, at least at the one-instanton level. 

\subsection{Sp(1) vs SU(2)}\label{a:su2}

Another fundamental example of Lie algebra isomorphism is $\mathfrak{sp}(1) \cong \mathfrak{su}(2)$. In SU(2) and Sp($N$) theories, an additional discrete parameter, the $\theta$-angle, can be incorporated. For these theories, the non-trivial values of the $\theta$-angle are restricted to $0$ and $\pi$, due to the presence of a discrete symmetry. The defect partition function of Sp($N$) is given by 
\begin{align}
    Z_{\rm defect}^{{\rm Sp}(N)_\theta}(x)&=\sum_{k=0}^\infty\mathfrak{q}^k\frac12\bigg[\frac1{|W({\rm O}(k)^+)|}\oint_{C_{\rm JK}}\left(\frac{d\phi_I}{2\pi i}\right)Z_{\mathrm{vec}}^{(k,+)}\cdot Z_{\mathrm{defect}}^{(k,+)}(x)\notag
    \\
    &\pm\frac1{|W({\rm O}(k)^-)|}\oint_{C_{\rm JK}}\left(\frac{d\phi_I}{2\pi i}\right)Z_{\mathrm{vec}}^{(k,-)}\cdot Z_{\mathrm{defect}}^{(k,-)}(x)\bigg],\label{Sp-integral}
\end{align}
where the sign $+$ corresponds to $\theta=0$ and $-$ corresponds to $\theta=\pi$. The explicit forms of 1-loop determinants of $Z_{\rm vec}$ and $Z_{\rm defect}$ are given in Appendix \ref{a:Sp}. Furthermore, the orders of the Weyl groups for the $O(l)_{\pm}$ components depend on the instanton number $l$ as follows: 
\begin{align}
    |W({\rm O}(2\ell)^+)|=&\frac{1}{2^{\ell-1}\ell!},\quad |W({\rm O}(2\ell{+}1)^+)|=\frac{1}{2^{\ell}\ell!},\cr |W({\rm O}(2\ell)^-)|=&\frac{1}{2^{\ell-1}(\ell{-}1)!},\quad |W({\rm O}(2\ell{+}1)^-)|=\frac{1}{2^{\ell}\ell!}.
\end{align}
Within the SU(2) gauge theory framework, the $\theta$-angle is realized by including a Chern-Simons-like contribution, $Z_{CS}^{(k)}=\prod_{I=1}^k e^{\phi_I \kappa}$, and the defect partition function is given by 
\begin{align}
     Z_{\rm defect}^{{\rm SU(2)}_\theta}(x)&=\sum_{k=0}^\infty\frac{\mathfrak{q}^k}{k!}\oint_{C_{\rm JK}}\left[\frac{d\phi_I}{2\pi i}\right]Z_{\mathrm{vec}}^{(k)}\cdot \left[Z_{CS}^{(k)}\right]^{\frac{\theta}{2\pi}}\cdot Z_{\mathrm{defect}}^{(k)}(x).
\end{align}
At the level of the fundamental qq-character associated to one-defect partition function, the isomorphism between $\mathfrak{sp}(1)$ and $\mathfrak{su}(2)$ has been confirmed in \cite{Haouzi:2020zls,Nawata:2023wnk}. The key point to establish the equivalence is to include an extra contribution in Sp(1) theory. In this section, we first review the results for one-defect partition function and then extend it to two-defect cases. 

In the case of one defect, the vacuum expectation value of the SU(2) qq-character equals that of the Sp(1) qq-character. However, the Sp(1) one-defect partition function itself differs from this expectation value by an extra instanton contribution, 
\begin{align}
    \frac{Z^{{\rm SU(2)}_\theta}_{\rm 1\ defect}(x)}{Z^{{\rm SU(2)}_\theta}_{\rm inst}}&=\langle\chi^{{{\rm SU(2)}_\theta}}(x)\rangle=\left\langle\chi^{{\rm Sp(1)}_\theta}(x)\right\rangle=\frac{Z_{\rm 1\ defect}^{{\rm Sp(1)}_\theta}(x)}{Z^{{\rm Sp(1)}_\theta}_{\rm inst}}+\mathfrak{q}Z_{\mathrm{extra}}^{{\rm Sp(1)}_\theta}(x),
    \label{ZSU21D}
\end{align}
with the extra term given by
\begin{align}
    Z_{\mathrm{extra}}^{{\rm Sp(1)}_{\theta=0}}(x)=-\frac{x\left(1+x^2\right)q_1q_2}{\left(x^2-q_1q_2\right)\left(x^2q_1q_2-1\right)},
\end{align}
and 
\begin{align}
    Z_{\mathrm{extra}}^{{\rm Sp(1)}_{\theta=\pi}}(x)=-\frac{\sqrt{q_1q_2}\left(1+q_1q_2\right)x^2}{\left(x^2-q_1q_2\right)\left(x^2q_1q_2-1\right)}.
\end{align}

Now present the results for the two-defect partition function, expanded to first order in the instanton counting parameter $\mathfrak{q}$:
\begin{align}
    \frac{Z^{{\rm SU(2)}_{\theta=0}}_{\rm 2\ defect}(x_1,x_2)}{Z^{{\rm SU(2)}_{\theta=0}}_{\rm inst}}&=\frac{(x_1-A)(x_2-A)(1-x_1A)(1-x_2A)}{x_1x_2A^2}\notag
    \\
    &+\mathfrak{q}\frac{q_1q_2}{x_1x_2(x_1-x_2q_1q_2)(x_2-x_1q_1q_2)(q_1q_2-A^2)(1-q_1q_2A^2)}\notag
    \\
    &\times\Big(\left(1+A^4\right)x_1x_2\left(-2x_1^2q_1q_2-2x_2^2q_1q_2+x_1x_2\left(1+q_1+q_2-2q_1q_2\right.\right.\notag
    \\
    &\left.\left.+q_1q_2^2+q_1^2q_2+q_1^2q_2^2\right)\right)+\left(A+A^3\right)\left(x_1+x_2+x_1x_2^2+x_1^2x_2\right)\left(x_1^2q_1q_2\right.\notag
    \\
    &\left.+x_2^2q_1q_2-x_1x_2-x_1x_2q_1^2q_2^2\right)+A^2x_1x_2\left(4x_1x_2\left(1+q_1^2q_2^2\right)+\left(x_1^2+x_2^2\right)\right.\notag
    \\
    &\cdot\left.\left(1-q_1-q_2-2 q_1 q_2-q_1^2 q_2-q_1 q_2^2+q_1^2 q_2^2\right)\right)\Big)+\mathcal{O}(\mathfrak{q}^2),
    \label{ZSU22D}
\end{align}
\begin{align}
    \frac{Z^{{{\rm Sp(1)}_{\theta=0}}}_{\rm 2\ defect}(x_1,x_2)}{Z^{{{\rm Sp(1)}_{\theta=0}}}_{\rm inst}}&=\frac{(x_1-A)(x_2-A)(1-x_1A)(1-x_2A)}{x_1x_2A^2}\notag
    \\
    &+\mathfrak{q}\frac{q_1 q_2 \left(x_1-A\right) \left(x_2-A\right) \left(1-x_1 A\right) \left(1-x_2 A\right)}{2 \left(1-q_1\right) \left(1-q_2\right) x_1 x_2 A}\notag
    \\
    &\times\left(\frac{\left(\sqrt{q_2}-\sqrt{q_1} x_1\right) \left(\sqrt{q_1}-\sqrt{q_2} x_1\right)
   \left(\sqrt{q_2}-\sqrt{q_1} x_2\right) \left(\sqrt{q_1}-\sqrt{q_2} x_2\right)}{\left(\sqrt{q_1 q_2}-x_1\right) \left(1-\sqrt{q_1 q_2}
   x_1\right) \left(\sqrt{q_1 q_2}-x_2\right) \left(1-\sqrt{q_1 q_2} x_2\right)}\right.\notag
   \\
   &\cdot\frac{1}{\left(A-\sqrt{q_1
   q_2}\right) \left(1-A \sqrt{q_1 q_2}\right) }+\frac{1}{\left(A+\sqrt{q_1
   q_2}\right) \left(1+A \sqrt{q_1 q_2}\right) }\notag
   \\
   &\cdot\frac{\left(\sqrt{q_2}+\sqrt{q_1} x_1\right) \left(\sqrt{q_1}+\sqrt{q_2} x_1\right)
   \left(\sqrt{q_2}+\sqrt{q_1} x_2\right) \left(\sqrt{q_1}+\sqrt{q_2} x_2\right)}{\left(\sqrt{q_1 q_2}+x_1\right) \left(1+\sqrt{q_1 q_2}
   x_1\right) \left(\sqrt{q_1 q_2}+x_2\right) \left(1+\sqrt{q_1 q_2} x_2\right)}\notag
   \\
   &+\left.\frac{2 \left(q_1 q_2+1\right)A}{\left(q_1 q_2-A^2\right) \left(1- q_1 q_2 A^2\right)}\right)+\mathcal{O}(\mathfrak{q}^2),
    \label{ZSP12D}
\end{align}
and 
\begin{align}
    \frac{Z^{{\rm SU(2)}_{\theta=\pi}}_{\rm 2\ defect}(x_1,x_2)}{Z^{{\rm SU(2)}_{\theta=\pi}}_{\rm inst}}&=\frac{(x_1-A)(x_2-A)(1-x_1A)(1-x_2A)}{x_1x_2A^2}\notag
    \\
    &+\mathfrak{q}\frac{\sqrt{q_1 q_2}}{x_1 x_2 \left(A^2-q_1 q_2\right) \left(A^2 q_1 q_2-1\right) \left(q_1 q_2 x_1-x_2\right) \left(q_1 q_2 x_2-x_1\right)}\notag
    \\
    &\times\Big(-\left(1+A^4\right)\left(q_1-1\right) \left(q_2-1\right) q_1 q_2 x_1 x_2 \left(x_1+x_2\right)\notag
    \\
   &+\left(A+A^3\right) \left(q_1+1\right) \left(q_2+1\right) x_1 x_2 \left(q_1 q_2
   x_1-x_2\right) \left(q_1 q_2 x_2-x_1\right)\notag
    \\
    &+A^2 \left(x_1+x_2\right)\left(q_1 q_2 \left(q_1 q_2+1\right) \left(x_1 x_2+1\right) \left(x_1^2+x_2^2\right)\right.\notag
   \\
   &-\left.\left(q_1^2 q_2^2+1\right) \left(\left(q_1 q_2+1\right) x_1^2
   x_2^2+\left(q_1+q_2\right) x_1 x_2\right)\right)\Big)+\mathcal{O}(\mathfrak{q}^2),
    \label{ZSU22Dpi}
\end{align}
\begin{align}
    \frac{Z^{{\rm Sp(1)}_{\theta=\pi}}_{\rm 2\ defect}(x_1,x_2)}{Z^{{\rm Sp(1)}_{\theta=\pi}}_{\rm inst}}&=\frac{(x_1-A)(x_2-A)(1-x_1A)(1-x_2A)}{x_1x_2A^2}\notag
    \\
    &+\mathfrak{q}\frac{q_1 q_2\left(A-x_1\right) \left(1-A x_1\right) \left(A-x_2\right) \left(1-A x_2\right)}{2 A \left(q_1-1\right) \left(q_2-1\right) x_1 x_2}\notag
    \\
    &\times\left(\frac{\left(\sqrt{q_1} x_1-\sqrt{q_2}\right) \left(\sqrt{q_1}-\sqrt{q_2} x_1\right) \left(\sqrt{q_1} x_2-\sqrt{q_2}\right) \left(\sqrt{q_1}-\sqrt{q_2}
   x_2\right)}{ \left(\sqrt{q_1 q_2}-x_1\right) \left(\sqrt{q_1 q_2} x_1-1\right) \left(\sqrt{q_1
   q_2}-x_2\right) \left(\sqrt{q_1 q_2} x_2-1\right)}\right.\notag
   \\
   &\cdot\frac{1}{\left(\sqrt{q_1 q_2}-A\right) \left(A \sqrt{q_1 q_2}-1\right)}-\frac{1}{\left(\sqrt{q_1 q_2}+A\right) \left(A \sqrt{q_1 q_2}+1\right)}\notag
   \\
   &\cdot\frac{\left(\sqrt{q_1} x_1+\sqrt{q_2}\right) \left(\sqrt{q_2} x_1+\sqrt{q_1}\right) \left(\sqrt{q_1} x_2+\sqrt{q_2}\right) \left(\sqrt{q_2}
   x_2+\sqrt{q_1}\right)}{\left(\sqrt{q_1 q_2}+x_1\right) \left(\sqrt{q_1 q_2} x_1+1\right) \left(\sqrt{q_1 q_2}+x_2\right) \left(\sqrt{q_1 q_2} x_2+1\right)}\notag
   \\
   &+\left.\frac{2 \left(A^2+1\right) \sqrt{q_1 q_2}}{\left(A^2-q_1 q_2\right) \left(A^2 q_1 q_2-1\right)}\right)+\mathcal{O}(\mathfrak{q}^2).
    \label{ZSP12Dpi}
\end{align}
The two-defect partition function can be decomposed into a factorized part, resembling the product of two one-defect partition function, plus a set of correlation terms. This structure motivates the following ansatz for its expectation value:
\begin{align}
     \left<\chi_{\rm 2\ defect}^{\theta=0}(x_1,x_2)\right> &:= \frac{Z^{{\rm SU(2)}_{\theta=0}}_{\rm 2\ defect}(x_1,x_2)}{Z^{{\rm SU(2)}_{\theta=0}}_{\rm inst}}=\frac{Z^{{\rm Sp(1)}_{\theta=0}}_{\rm 2\ defect}(x_1,x_2)}{Z^{{\rm Sp(1)}_{\theta=0}}_{\rm inst}}\notag
    \\
    &+\mathfrak{q}Z_{\rm extra}^{{\rm Sp(1)}_{\theta=0}}(x_1)\frac{Z^{{\rm Sp(1)}_{\theta=0}}_{\rm 1\ defect}(x_2)}{Z^{{\rm Sp(1)}_{\theta=0}}_{\rm inst}}+\mathfrak{q}Z_{\rm extra}^{{\rm Sp(1)}_{\theta=0}}(x_2)\frac{Z^{{\rm Sp(1)}_{\theta=0}}_{\rm 1\ defect}(x_1)}{Z^{{\rm Sp(1)}_{\theta=0}}_{\rm inst}}\notag
    \\
    &+\mathfrak{q}^2Z_{\rm extra}^{{\rm Sp(1)}_{\theta=0}}(x_1)Z_{\rm extra}^{{\rm Sp(1)}_{\theta=0}}(x_2)+\mathfrak{q}C_0^{\theta=0}\cdot W_{\mathbf{2}}+\mathfrak{q} C_1^{\theta=0}+\mathfrak{q}^2 C_2^{\theta=0},
\end{align}
where the coefficients are found to be 
\begin{align}
C_0^{\theta=0}&=\frac{x_1x_2(x_1+x_2)(1+x_1x_2)q_1q_2(1-q_1)(1-q_2)(1+q_1q_2)}{(x_1^2-q_1q_2)(x_2^2-q_1q_2)(1-x_1^2q_1q_2)(1-x_2^2q_1q_2)},
\end{align}
\begin{align}
    C_1^{\theta=0}&=\frac{q_1q_2(1-q_1)(1-q_2)(1+q_1q_2)}{(x_1^2-q_1q_2)(x_2^2-q_1q_2)(x_1-x_2q_1q_2)(x_2-x_1q_1q_2)(1-x_1^2q_1q_2)(1-x_2^2q_1q_2)}\notag
    \\
    &\times\Big(-x_1^2x_2^2(1+2x_1x_2+2x_1^2+2x_2^2+x_1^2x_2^2)+(-x_1x_2+x_1^4+x_2^4+x_1x_2^3+x_1^3x_2\notag
    \\
    &+2x_1^2x_2^2+2x_1^2x_2^4+2x_1^4x_2^2+2x_1^3x_2^3+x_1^2x_2^6+x_1^6x_2^2+x_1^3x_2^5+x_1^5x_2^3+2x_1^4x_2^4-x_1^5x_2^5)q_1q_2\notag
    \\
    &-x_1^2x_2^2(1+2x_1x_2+2x_1^2+2x_2^2+x_1^2x_2^2)q_1^2q_2^2\Big),
    \\
    C_2^{\theta=0}&=\frac{x_1^2x_2^2q_1^2q_2^2(1-q_1)(1-q_2)(1+q_1q_2)}{(x_1x_2-q_1q_2)(x_1-x_2q_1q_2)(x_2-x_1q_1q_2)(-1+x_1x_2q_1q_2)}\notag
    \\
    &\times\frac{1}{(q_1q_2-x_1^2)(q_1q_2-x_2^2)(1-x_1^2q_1q_2)(1-x_2^2q_1q_2)}\notag
    \\
    &\times\Big(x_1x_2(1+x_1x_2+x_1^2+x_2^2+x_1^2x_2^2)+x_1x_2(1+x_1x_2+x_1^2+x_2^2+x_1^2x_2^2)q_1^2q_2^2\notag
    \\
    &-(x_1^2+x_2^2+x_1x_2+x_1x_2^3+x_1^3x_2+2x_1^2x_2^2+x_1^2x_2^4+x_1^4x_2^2+x_1^3x_2^3)q_1q_2\Big).
\end{align}
In the case of $\theta=\pi$, we found a similar expression, 
\begin{align}
     \left<\chi_{\rm 2\ defect}^{\theta=\pi}(x_1,x_2)\right> &:= \frac{Z^{{\rm SU(2)}_{\theta=\pi}}_{\rm 2\ defect}(x_1,x_2)}{Z^{{\rm SU(2)}_{\theta=\pi}}_{\rm inst}}=\frac{Z^{{\rm Sp(1)}_{\theta=\pi}}_{\rm 2\ defect}(x_1,x_2)}{Z^{{\rm Sp(1)}_{\theta=\pi}}_{\rm inst}}\notag
    \\
    &+\mathfrak{q}Z_{\rm extra}^{{\rm Sp(1)}_{\theta=\pi}}(x_1)\frac{Z^{{\rm Sp(1)}_{\theta=\pi}}_{1\,defect}(x_2)}{Z^{{\rm Sp(1)}_{\theta=\pi}}_{\rm inst}}+\mathfrak{q}Z_{\rm extra}^{{\rm Sp(1)}_{\theta=\pi}}(x_2)\frac{Z^{{\rm Sp(1)}_{\theta=\pi}}_{\rm 1\ defect}(x_1)}{Z^{{\rm Sp(1)}_{\theta=\pi}}_{\rm inst}}\notag
    \\
    &+\mathfrak{q}^2Z_{\rm extra}^{{\rm Sp(1)}_{\theta=\pi}}(x_1)Z_{\rm extra}^{{\rm Sp(1)}_{\theta=\pi}}(x_2)+\mathfrak{q}C_0^{\theta=\pi}\cdot W_{\mathbf{2}}+\mathfrak{q} C_1^{\theta=\pi}+\mathfrak{q}^2 C_2^{\theta=\pi},
\end{align}
with different coefficients, 
\begin{align}
    C_0^{\theta=\pi}&=\frac{\left(1-q_1\right) \left(1-q_2\right) \sqrt{q_1 q_2} x_1 x_2 \left(\left(1+q_1 q_2\right)^2 x_1 x_2+q_1 q_2 \left(x_1^2+1\right)\left(x_2^2+1\right)\right)}{\left(x_1^2-q_1 q_2\right) \left(x_2^2-q_1 q_2\right) \left(q_1 q_2 x_1^2-1\right) \left(q_1 q_2 x_2^2-1\right)}
    \\
    C_1^{\theta=\pi}&=-\frac{\left(1-q_1\right) \left(1-q_2\right) \sqrt{q_1 q_2} \left(x_1+x_2\right)}{\left(x_1^2-q_1 q_2\right) \left(x_2^2-q_1 q_2\right) \left(q_1 q_2 x_1^2-1\right) \left(q_1 q_2 x_2^2-1\right) \left(q_1 q_2 x_1-x_2\right) \left(q_1 q_2 x_2-x_1\right) }\notag
    \\
    &\times\Big(x_1^2 x_2^2\left(1+q_1^4q_2^4\right)+q_1q_2\left(x_1^4 x_2^4+x_1^3 x_2^3+x_1 x_2^3+x_1^3 x_2+x_1^2 x_2^2+x_1 x_2-x_1^2-x_2^2\right)\notag
    \\
    &+q_1^2q_2^2\left(1-x_1^2-x_2^2-x_1 x_2^3-x_1^3 x_2 -x_1 x_2^5 -x_1^5 x_2 -x_1^2x_2^4 -x_1^4 x_2^2-2 x_1^3 x_2^3 - x_1^3x_2^5\right.\notag
    \\
    &\left.-x_1^5x_2^3 +x_1^4 x_2^4\right)+q_1^3 q_2^3\left(-x_1^2-x_2^2+x_1 x_2+x_1 x_2^3+x_1^3 x_2+x_1^2 x_2^2+x_1^3 x_2^3+x_1^4 x_2^4\right)\Big)
    \\
    C_2^{\theta=\pi}&=\frac{\left(1-q_1\right)\left(1-q_2\right)\left(1+q_1 q_2\right) q_1 q_2 x_1^3 x_2^3 }{\left(q_1 q_2-x_1^2\right) \left(q_1 q_2-x_2^2\right) \left(q_1 q_2 x_1^2-1\right)\left(q_1 q_2 x_2^2-1\right)}\notag
    \\
    &\times\frac{\left(q_1^4 q_2^4 x_1 x_2+q_1^3 q_2^3 x_1 x_2-q_1^2 q_2^2 \left(x_1^2+1\right) \left(x_2^2+1\right)+q_1 q_2 x_1 x_2+x_1 x_2\right)}{ \left(q_1 q_2 x_1-x_2\right) \left(q_1 q_2 x_2-x_1\right)  \left(q_1 q_2-x_1 x_2\right) \left(q_1 q_2 x_1 x_2-1\right)}.
\end{align}

\section{Derivation for \eqref{eq:Zhatc1c2} and $d_i$}\label{a:proof}

In this appendix, we demonstrate that if we choose $\omega^{\vee}=0$ and only insert the Wilson loop at the south pole with $\mathbf{r}_{\rm S}=\mathbf{r}_i$, the blowup equation has the expression \eqref{eq:Zhatc1c2} for $b=b_0-2,b_0,b_0+2$. The demonstration is highly reliant on the constants $(d_1,\cdots,d_r)$ for the simple Lie algebra $G$.

For the 5d pure gauge theory engineered from the corresponding Calabi-Yau threefold $X$, the one-loop contribution is from the positive roots of $G$. For any positive root, it can be expanded in the simple-root basis ${\alpha_i}$ with non-negative integer coefficients.
This suggests that the K\"ahler parameters for $X$ are 
\begin{align}
    t_i=\alpha_i\cdot a,\,\,i=1,\cdots,r,\qquad t_{r+1}=\log{\frak{q}}+\sum_{i=1}^r d_it_i,
\end{align}
where $d_i$ are some possible shifts to the base K\"ahler parameter.

K\"ahler parameters parametrize the K\"ahler moduli space of the Calabi-Yau threefold $X$. By definition, if we expand the gauge theory BPS partition function in terms of $Q_i=e^{-t_i}$, the degrees of $Q_i$ are always non-negative integers. From the logic that the partition function can be solved from three different fluxes $b$, the constants $d_i$ are expected to be calculated from the blowup equation and the prepotential. 

The prepotential describes the low-energy dynamics for the 5d gauge theory. For a pure super Yang-Mills theory, it is written as \cite{intriligator:1997pq}
\begin{align}
    \mathcal{F}=\frac{1}{6}\sum_{\alpha\in\Delta^+} (\alpha\cdot a)^3+\log{\frak q}\,\frac{1}{4h_G^{\vee}} \sum_{\alpha\in\Delta^+} (\alpha\cdot a)^2+\cdots,
\end{align}
where $\Delta^+$ is the set of positive roots for the gauge algebra and $\cdots$ are irrelevant terms for our discussion. By substituting this prepotential into the blowup equation, we can reorganize the classical factor in the flux sum as 
\begin{align}
    \exp\left({\sum_{i=1}^{r}f_i(\vec{n},b) t_i+\frac{\vec{n}\cdot \vec{n}}{2}\log\frak q}\right),
\end{align}
where 
\begin{align}
    \sum_{i=1}^{r}f_i(\vec{n},b) t_i=\sum_{\alpha\in\Delta^{+}}\left[\frac{1}{2}(\alpha\cdot \vec{n})^2 +\frac{b}{2h_G^{\vee}}(\alpha\cdot \vec{n})\right](\alpha\cdot a),
\end{align}

In the bootstrap process, three selected fluxes $b$ are used to solve the instanton partition function recursively. Schematically, if the absolute value of $b$ is larger, the minima of $f_i$ will be smaller. The constants $d_i$ are chosen to be the maximal shift of $\log\frak q$, so that $d_i$ can be solved from the minimal values of $f_i(\vec{n},b)$ for $b=b_0-2,b_0,b_0+2$, where $b_0\in\{0,1\}$ is the parity of $h_G^\vee$.  At the one-instanton level, the integer lattice that satisfies $\vec{n}\cdot \vec{n}=2$ contributes to the flux sum. 
Define
\begin{align}
    g_{i,b}=\min (\{f_i(\vec{n},b)|\vec{n}\cdot\vec{n}=2\}),\quad i=1,\cdots,r,
\end{align}
from the reasons above, $d_i$ can be computed as:
\begin{align}
    d_i=\min(g_{i,b_0-2},g_{i,b_0},g_{i,b_0+2}),
\end{align}

If we insert a Wilson loop operator in the representation ${\bf r}$ at the south pole of the blown-up space, the minimal value of $f_{i,b}$ will have a constant shift due to the highest weight of $\bf r$. Then it is straightforward to see that if the highest weight is smaller than $(d_1,\cdots,d_r)$, then $c_2=0$. If it is equal to $(d_1,\cdots,d_r)$, then the classical factor in the flux sum will contribute another term, which gives $c_2\neq 0$. 

\newpage
\section{Wilson loops in antisymmetric representations}\label{a:su(n)-k}

Hereby we list the one-instanton contribution to the BPS sector of SU($N$) Wilson loops (with Coulomb branch parameters turned off) in antisymmetric representation. The Chern-Simons level for the following theories are $\kappa=\frac{1}{2}(N-4+N_{\mathbf{F}}-N_{\overline{\mathbf{F}}})$. We introduce the notation, $\bar{\cal F}_{\bf r}:={\cal I}^{1-|q|}{\cal F}^{(1)}_{\bf r}$. Then we have, 
\begin{align}
    &\bar{\cal F}_{\rm SU(4)+asym}= -\frac{2 v^4 (3+14 v^2+3 v^4)}{(1-v^2)^6},\quad \bar{\cal F}_{{\rm SU(4)+asym+1F}}= \frac{20 v^{5} (1+v^2)}{(1-v^2)^6},\cr
    &\bar{\cal F}_{{\rm SU(4)+asym+2F}}=\frac{v^2 (1-6 v^2-30 v^4-6 v^6+v^8)}{(1-v^2)^6},\cr
    &\bar{\cal F}_{{\rm SU(4)+asym+3F}}=-\frac{(4 v^{3} (1-6 v^2-6 v^4+v^6)}{(1-v^2)^6},\cr
    &\bar{\cal F}_{{\rm SU(4)+asym+1F+1{\bar F}}}=-\frac{2 v^4 (3+14 v^2+3 v^4)}{(1-v^2)^6},\quad \bar{\cal F}_{{\rm SU(4)+asym+1F+2{\bar F}}}= \frac{20 v^{5} (1+v^2)}{(1-v^2)^6},\cr
    &\bar{\cal F}_{{\rm SU(4)+asym+1F+3{\bar F}}}=\frac{v^2 (1-6 v^2-30 v^4-6 v^6+v^8)}{(1-v^2)^6},\cr
    &\bar{\cal F}_{{\rm SU(4)+asym+2F+2{\bar F}}}= -\frac{2 v^4 (3+14 v^2+3 v^4)}{(1-v^2)^6},\quad \bar{\cal F}_{{\rm SU(4)+asym+2F+3{\bar F}}}=\frac{20 v^{5} (1+v^2)}{(1-v^2)^6},\cr
    &\bar{\cal F}_{{\rm SU(4)+asym+3F+3{\bar F}}}=-\frac{2 v^4 (3+14 v^2+3 v^4)}{(1-v^2)^6}.
\end{align}
Here we use the label ${\rm SU}(N)+n\ {\rm asym}+N_f{\rm F}+N_{\bar{f}}\bar{F}$ to indicate the gauge group $G={\rm SU}(N)$, the number of antisymmetric hypermultiplets $n$, the number of fundamental matters $N_f$, and the number of anti-fundamental ones $N_{\bar{f}}$. 
We note that the expressions are symmetric about the exchange of fundamental and anti-fundamental representations in SU(4) case, since the antisymmetric representation here is real. 
\begin{align}
    &\bar{\cal F}_{\rm SU(5)+asym}=-\frac{5 v^{5} (2+19 v^2+19 v^4+2 v^6)}{(1-v^2)^8},\quad \bar{\cal F}_{{\rm SU(5)+asym+1F}}= \frac{15 v^6 (3+8 v^2+3 v^4)}{(1-v^2)^8},\cr
    &\bar{\cal F}_{{\rm SU(5)+asym+2F}}=\frac{v^{3} (1-8 v^2-98 v^4-98 v^6-8 v^8+v^{10})}{(1-v^2)^8},\cr
    &\bar{\cal F}_{{\rm SU(5)+asym+3F}}= -\frac{5 v^4 (1-8 v^2-28 v^4-8 v^6+v^8)}{(1-v^2)^8},\quad \bar{\cal F}_{{\rm SU(5)+asym+4F}}=\frac{15 v^{5} (1-8 v^2-8 v^4+v^6)}{(1-v^2)^8},\cr
    &\bar{\cal F}_{{\rm SU(5)+asym+1F+1{\bar F}}}=-\frac{5 v^{5} (2+19 v^2+19 v^4+2 v^6)}{(1-v^2)^8},\quad \bar{\cal F}_{{\rm SU(5)+asym+1F+2{\bar F}}}= \frac{15 v^6 (3+8 v^2+3 v^4)}{(1-v^2)^8},\nonumber
\end{align}
\begin{align}
    &\bar{\cal F}_{{\rm SU(5)+asym+1F+3{\bar F}}}=\frac{v^{3} (1-8 v^2-98 v^4-98 v^6-8 v^8+v^{10})}{(1-v^2)^8)},\cr
    &\bar{\cal F}_{{\rm SU(5)+asym+1F+4{\bar F}}}= -\frac{5 v^4 (1-8 v^2-28 v^4-8 v^6+v^8)}{(1-v^2)^8},\quad \bar{\cal F}_{{\rm SU(5)+asym+2F+1{\bar F}}}=\frac{10 v^6 (4+13 v^2+4 v^4)}{(1-v^2)^8},\cr
    &\bar{\cal F}_{{\rm SU(5)+asym+2F+2{\bar F}}}= -\frac{5 v^{5} (2+19 v^2+19 v^4+2 v^6)}{(1-v^2)^8},\quad \bar{\cal F}_{{\rm SU(5)+asym+2F+3{\bar F}}}=\frac{15 v^6 (3+8 v^2+3 v^4)}{(1-v^2)^8},\cr
    &\bar{\cal F}_{{\rm SU(5)+asym+2F+4{\bar F}}}=\frac{v^{3} (1-8 v^2-98 v^4-98 v^6-8 v^8+v^{10})}{(1-v^2)^8},\nonumber
\end{align}
\begin{align}
    &\bar{\cal F}_{{\rm SU(5)+asym+3F+1{\bar F}}}=-\frac{105 v^{7} (1+v^2)}{(1-v^2)^8},\quad \bar{\cal F}_{{\rm SU(5)+asym+3F+2{\bar F}}}=\frac{10 v^6 (4+13 v^2+4 v^4)}{(1-v^2)^8},\cr
    &\bar{\cal F}_{{\rm SU(5)+asym+3F+3{\bar F}}}= -\frac{5 v^{5} (2+19 v^2+19 v^4+2 v^6)}{(1-v^2)^8},\quad \bar{\cal F}_{{\rm SU(5)+asym+3F+4{\bar F}}}=\frac{15 v^6 (3+8 v^2+3 v^4)}{(1-v^2)^8},\cr
    &\bar{\cal F}_{{\rm SU(5)+asym+4F+1{\bar F}}}=\frac{v^2 (1-8 v^2+28 v^4+168 v^6+28 v^8-8 v^{10}+v^{12})}{(1-v^2)^8},\cr
    &\bar{\cal F}_{{\rm SU(5)+asym+4F+2{\bar F}}}=-\frac{105 v^{7} (1+v^2)}{(1-v^2)^8},\quad \bar{\cal F}_{{\rm SU(5)+asym+4F+3{\bar F}}}=\frac{10 v^6 (4+13 v^2+4 v^4)}{(1-v^2)^8},\cr
    &\bar{\cal F}_{{\rm SU(5)+asym+4F+4{\bar F}}}=-\frac{5 v^{5} (2+19 v^2+19 v^4+2 v^6)}{(1-v^2)^8}.
\end{align}
We observe an interesting equality in this special limit with {\it Coulomb branch parameters turned off}, 
\begin{equation}
    \bar{\cal F}_{{\rm SU}(N)_{\kappa}{\rm +asym}+N_f {\rm F}+N_{\bar f}{\rm \bar{F}}}=\bar{\cal F}_{{\rm SU}(N)_{\kappa}{\rm +asym}+(N-N_{\bar f}) {\rm F}+(N-N_f){\rm \bar{F}}},
\end{equation}
where $\kappa=\frac{1}{2}(N-4+N_{\mathbf{F}}-N_{\overline{\mathbf{F}}})$. 
We checked it also for $N=6,7$, but we do not have any physical explanation for this equality.

\clearpage
\bibliographystyle{JHEP}     
{\small{\bibliography{reference}}}

\end{document}